\documentclass[journal]{IEEEtran}

\usepackage{cite}
\usepackage{booktabs}
\usepackage{multirow}
\usepackage[caption=false]{subfig}
\usepackage{amsmath}
\usepackage{verbatim}

\usepackage{xcolor}

\DeclareMathOperator*{\argmin}{argmin}

\ifCLASSINFOpdf
  \usepackage[pdftex]{graphicx}

  \graphicspath{{./png/}}

\else

\fi

\usepackage[ruled,vlined]{algorithm2e}
\SetKwInput{KwInput}{Input}                
\SetKwInput{KwOutput}{Output}                
\SetKwInput{KwReturn}{return}              

\begin{document}

\title{Enhanced Standard Compatible Image Compression Framework based on Auxiliary Codec Networks}

\author{Hanbin~Son, Taeoh~Kim, Hyeongmin~Lee,
        and~Sangyoun~Lee,~\IEEEmembership{Member,~IEEE}
\thanks{This work was partly supported by the Institute of Information \& communications Technology Planning \& Evaluation (IITP) grant funded by the Korea government, Ministry of Science and ICT (MSIT) (No. 2021-0-00172, The development of human Re-identification and masked face recognition based on CCTV camera) and the Institute of Information \& communications Technology Planning \& Evaluation (IITP) grant funded by the Korea government, Ministry of Science and ICT (MSIT) (No.2016-0-00197, Development of the high-precision natural 3D view generation technology using smart-car multi sensors and deep learning).}
\thanks{H. Son, T. Kim, H. Lee and S. Lee are with the School of Electrical and Electronic Engineering, Yonsei University, Seoul, South Korea (e-mail: hbson@yonsei.ac.kr; kto@yonsei.ac.kr; minimonia@yonsei.ac.kr; syleee@yonesi.ac.kr)}
\thanks{Corresponding author: Sangyoun Lee}}

\markboth{IEEE Transactions on Image Processing, Vol. XX, No. X, X XXXX}%
{Shell \MakeLowercase{\textit{et al.}}: Bare Demo of IEEEtran.cls for IEEE Journals}

\maketitle

\begin{abstract}
Recent deep neural network-based research to enhance image compression performance can be divided into three categories: learnable codecs, postprocessing networks, and compact representation networks. The learnable codec has been designed for end-to-end learning beyond the conventional compression modules. The postprocessing network increases the quality of decoded images using example-based learning. The compact representation network is learned to reduce the capacity of an input image, reducing the bit rate while maintaining the quality of the decoded image. However, these approaches are not compatible with existing codecs or are not optimal for increasing coding efficiency. Specifically, it is difficult to achieve optimal learning in previous studies using a compact representation network due to the inaccurate consideration of the codecs. In this paper, we propose a novel standard compatible image compression framework based on auxiliary codec networks (ACNs). In addition, ACNs are designed to imitate image degradation operations of the existing codec, which delivers more accurate gradients to the compact representation network. Therefore, compact representation and postprocessing networks can be learned effectively and optimally. We demonstrate that the proposed framework based on the JPEG and High Efficiency Video Coding standard substantially outperforms existing image compression algorithms in a standard compatible manner.

\end{abstract}

\begin{IEEEkeywords}
Image compression, deep neural networks, compact representation, JPEG, High Efficiency Video Coding.
\end{IEEEkeywords}

\IEEEpeerreviewmaketitle

\section{Introduction}
\IEEEPARstart{W}{ith} the development of media technology, the demand for live streaming or communicating using high-resolution visual data has increased, requiring better performance of image and video compression algorithms. Standard algorithms have been carefully developed and released for compatibility between the encoder and decoder of compression algorithms across users.

The JPEG standard~\cite{wallace1992jpeg}, a traditional image compression algorithm, has been the most widely used in still-image compression because of its simplicity and compatibility. Its block partitioning, transform, quantization, and entropy-coding-based scheme widely affects many other image and video compression standards, such as JPEG2000~\cite{rabbani2002jpeg2000}, H.264/AVC~\cite{wiegand2003overview}, and High Efficiency Video Coding (HEVC)~\cite{sullivan2012overview}. Recent video coding standards~\cite{wiegand2003overview, sullivan2012overview} have adopted prediction-based coding methods to reduce the spatial and temporal redundancy of input video. Prediction-based coding increases the complexity of the compression algorithm but produces much better compression performance.

\begin{figure*}[!th]
	\centering
	\subfloat[Learnable codecs]
	{\includegraphics[width=0.497\linewidth]{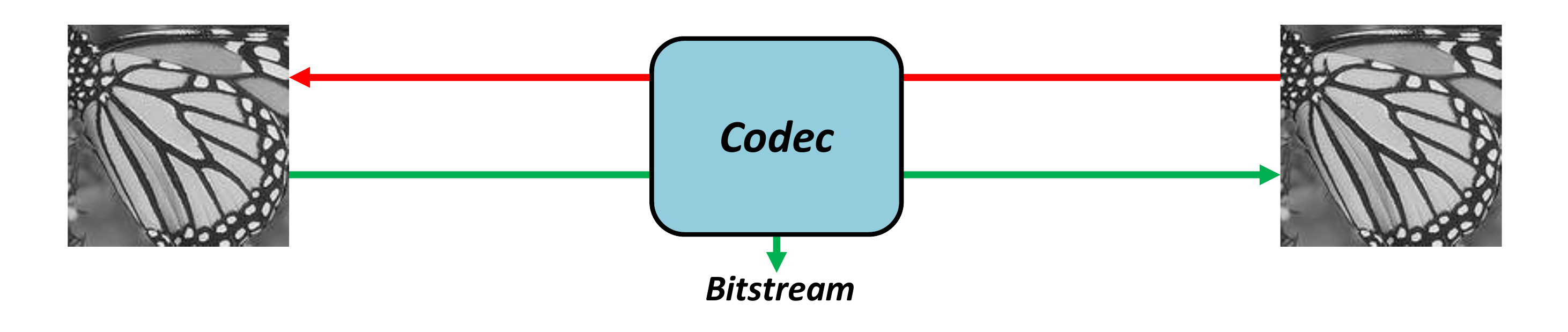}}\
	\hfill
	\subfloat[Postprocessing networks (PPNet)]
	{\includegraphics[width=0.497\linewidth]{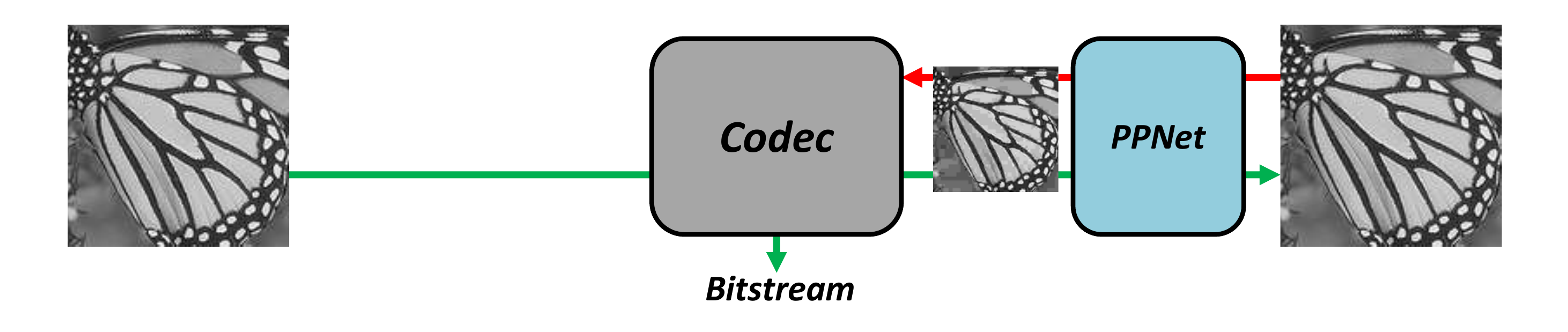}}\ \\
	\subfloat[Standard compatible frameworks based on the compact representation network (CRNet)]
	{\includegraphics[width=0.497\linewidth]{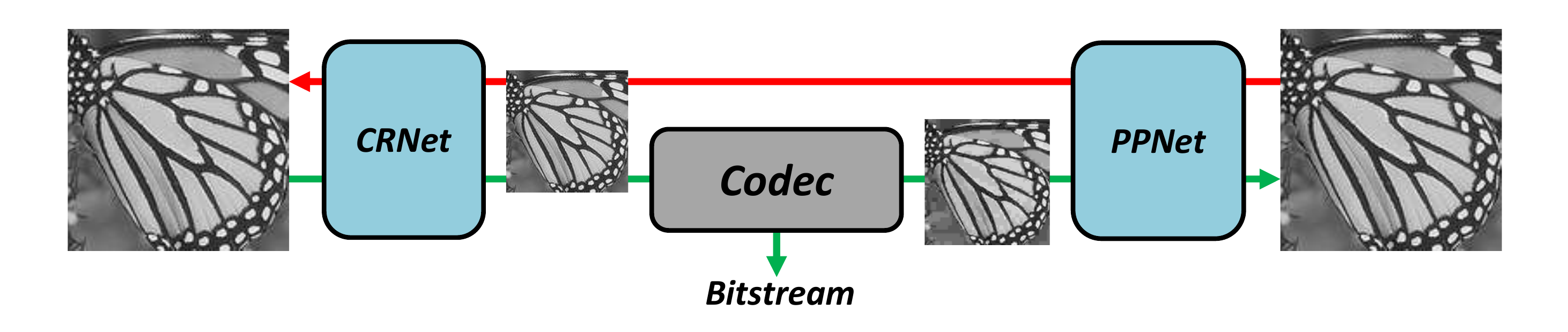}}\
	\hfill
	\subfloat[Proposed framework based on the auxiliary codec network (ACN)]
	{\includegraphics[width=0.497\linewidth]{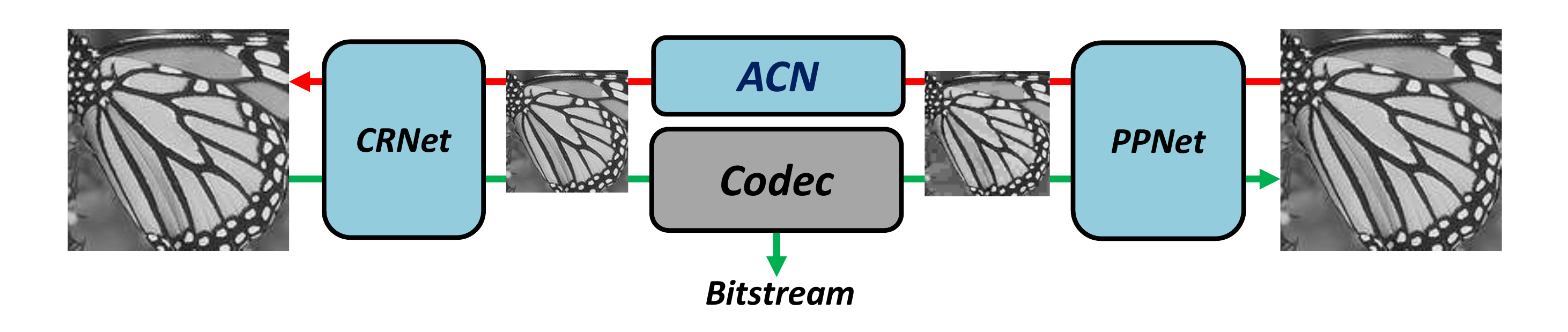}}\ \\
	\caption{Conceptual comparison between frameworks. Green and red arrows indicate forward (or inference) and backward pass (or gradients) to train the CRNet, respectively. Gray modules indicate that it is not differentiable or a standard codec. Blue modules indicate a differentiable network.}
	\label{fig_frameworkcomparison}
\end{figure*}

On the other hand, compression frameworks with end-to-end trainable deep neural networks~\cite{toderici2015variable, toderici2017full, theis2017lossy, balle2016end, balle2018variational, johnston2018improved, li2018learning2, mentzer2018conditional, minnen2018joint, lee2018context} (learnable codecs in this paper) have been proposed based on the rapid development of deep learning.
The approaches use trainable networks to produce bitstreams and reconstruct the original image (Fig.~\ref{fig_frameworkcomparison} (a)). Although these kinds of approaches structurally consider the compression ratio and reconstruction quality, their performance is still undesirable and incompatible with standard codecs, which decreases the algorithm's utility.

It is easy to propose a method to restore an image after the compression process to improve the compression performance while being compatible with standard codecs. Following the developments of convolutional neural networks (CNN), such as ResNet~\cite{he2016deep}, DenseNet~\cite{huang2017densely}, and attention networks~\cite{hu2018squeeze, wang2018non}, the CNN-based image postprocessing algorithms~\cite{kim2016accurate, ledig2017photo, tong2017image, liu2018non, cavigelli2017cas, galteri2017deep, tai2017memnet, wang2016d3, lim2017enhanced, zhang2018image} have drastically improved the performance of image restoration. These kinds of approaches are designated as a postprocessing network (PPNet) in this paper. Although PPNets perform well in reconstruction and are compatible with standard codecs, they only efficiently increase the visual quality of the reconstructed image but do not consider the compression ratio (Fig.~\ref{fig_frameworkcomparison} (b)).

The preprocessing and postprocessing-based coding strategies have been applied to consider both image quality and compression ratio. These kinds of approaches place a spatially downsampled image into the codec and upsample~\cite{bruckstein2003down, lin2006adaptive, wu2009low}, or are processed by the PPNet~\cite{afonso2018video, li2018convolutional}. Generally, reducing the spatial size of an image can increase the compression ratio.
However, the approaches only work at a low bit-rate setting~\cite{bruckstein2003down, lin2006adaptive, wu2009low}, and the predefined downsampling operations degrade detailed information and increase the ratio of high-frequency components, causing an increment in the bitstream.

The content-adaptive downsampling algorithm~\cite{zhang2011interpolation} or learning-based downsampling algorithms~\cite{kim2018task, li2018learning, sun2020learned, jiang2017end, zhao2019learning} have been proposed to overcome these problems. These algorithms train the content-adaptive downsampling (called compact representation in~\cite{jiang2017end}, abbreviated to CRNet in this paper) operation using the reconstruction loss after the PPNet. These approaches can achieve both a high compression ratio and better reconstruction quality with two networks. However, backward gradients from the loss function for the CRNet do not consider the degradation process through the standard codec (Fig.~\ref{fig_frameworkcomparison} (c)) because the standard codec, including the quantization process, is a nondifferentiable module.

In this paper, we propose a novel standard compatible end-to-end image compression framework based on auxiliary codec networks (ACNs). The ACNs are designed to imitate the forward image degradation process of existing codecs in differentiable networks to provide the correct backward gradients for training the CRNet (Fig.~\ref{fig_frameworkcomparison} (d)). These gradients allow the compact represented image to consider both the degradation process by the ACN and the reconstruction process by the PPNet.
Based on ACNs, both the CRNet and PPNet are learned together to achieve better image compression performance in a standard compatible manner. 
In addition, a bit estimation network (BENet) is proposed for training as a regularization function to prevent undesired bit-rate increments. As recent CRNet-based~\cite{kim2018task, li2018learning, sun2020learned, jiang2017end, zhao2019learning} generate models at a single level, the proposed framework is also trained and optimized per codec and rate.

The contributions of this paper are summarized as follows:
\begin{itemize}
\item We propose a novel CNN architecture called the ACN, based on the prior of the image compression process to effectively and precisely train the CRNet.

\item Based on the ACN, we propose an enhanced compression framework based on the collaborative learning scheme between the ACN, PPNet, and CRNet. Furthermore, the BENet facilitates training using a proper bit prediction, preventing undesirable artifacts in compactly represented images.

\item The framework is compatible with compression algorithms from the standard codecs to learning-based codecs and any off-the-shelf image restoration networks. Based on the highly accurate ACNs for two standard codecs: JPEG and HEVC, our framework exhibits state-of-the-art results compared to other image compression algorithms, including standards and learnable codecs.

\end{itemize}

\section{Related Work}
\subsection{Compression Frameworks Based on End-to-end Trainable Networks (Learnable Codecs)}
As deep learning has been successful in the field of image processing, Toderici \emph{et~al.} \cite{toderici2015variable, toderici2017full} first proposed an end-to-end deep neural network-based approach in image compression. An input image with dimensions reduced through an auto-encoder is stored as a binary vector for a given compression rate and is optimized for minimum distortion.

As the possibility of the strong modeling capacity of a neural network is revealed, many follow-up studies have been conducted.
Theis~\emph{et~al.}~\cite{theis2017lossy} proposed a compressive auto-encoder based on a residual neural network~\cite{he2016deep} and used a Laplace-smoothed histogram as the entropy model. Ball{\'e} \emph{et~al.} \cite{balle2016end} jointly optimized the entire model for rate-distortion performance using a generalized divisive normalization transform. Further, Ball{\'e} \emph{et~al.} \cite{balle2018variational} proposed a hyperprior to effectively capture spatial redundancy in the latent encoding. In addition, Johnston \emph{et~al.} \cite{johnston2018improved} proposed a priming technique and spatially adaptive entropy model for image compression. Moreover, Li \emph{et~al.} \cite{li2018learning2} proposed a content-weighted method based on spatially adaptive importance map learning. Mentzer \emph{et~al.} \cite{mentzer2018conditional} proposed a model that concurrently trains a context model with an encoder and used three-dimensional convolutional networks. Minnen \emph{et~al.} \cite{minnen2018joint} and Lee \emph{et~al.} \cite{lee2018context} combined a context-adaptive entropy model and hyperprior, producing substantial performance improvements.

The primary difficulties of deep image compression algorithms include making the nondifferential quantization process end-to-end trainable, designing an entropy model that predicts the bitstream generated from coefficients, and enabling compression considering both the bit rate and distortion. However, although many deep network-based approach algorithms have been developed, it is challenging to replace conventional compression schemes due to compatibility.
Furthermore, although the state-of-the-art approaches outperform even the Better Portable Graphics (BPG)~\cite{bellard2015bpg} codec, which is designed based on the intra-mode of HEVC, a significant performance improvement has not been demonstrated.

\subsection{Postprocessing Networks}
Following the success of the deep learning-based approach for high-level vision problems, methods for low-level vision problems, such as image super-resolution and compression artifact removal, have improved progressively. In single-image super-resolution, Dong \emph{et~al.}~\cite{dong2015image} first proposed a CNN called the super-resolution CNN (SRCNN) for the super-resolution problem to learn end-to-end mapping from downsampled images to high-resolution images. In addition, Kim \emph{et~al.} \cite{kim2016accurate} proposed a deeper network architecture with the residual skip connection. Ledig \emph{et~al.}~\cite{ledig2017photo} and Lim \emph{et~al.}~\cite{lim2017enhanced} proposed networks based on ResNet~\cite{he2016deep}, and Tong \emph{et~al.}~\cite{tong2017image} proposed a network based on DenseNet~\cite{huang2017densely}. Based on the attention networks~\cite{wang2018non, hu2018squeeze} in the recognition area, Liu \emph{et~al.}~\cite{liu2018non} and Zhang \emph{et~al.}~\cite{zhang2018image} proposed attention-based restoration networks.

For the removal of compression artifacts, Dong \emph{et~al.}~\cite{dong2015compression} proposed a network that is slightly deeper than SRCNN~\cite{dong2015image} to reduce artifacts in the intermediate feature maps. Like super-resolution networks, deeper networks~\cite{cavigelli2017cas, galteri2017deep, tai2017memnet} have been proposed. Some researchers have proposed frequency-based networks~\cite{wang2016d3, zhang2018dmcnn, zheng2019implicit, kim2019sf} to restore images from frequency transform-based compression algorithms (\textit{e.g.,} JPEG). Although these approaches can be used to recover a decoded image after compression algorithms or recover the original resolution if an image is downsampled before compression, they can only treat already compressed images and are not accessible to the bit-rate-related module.

\begin{figure*}
\centering
\includegraphics[width=0.8\linewidth]{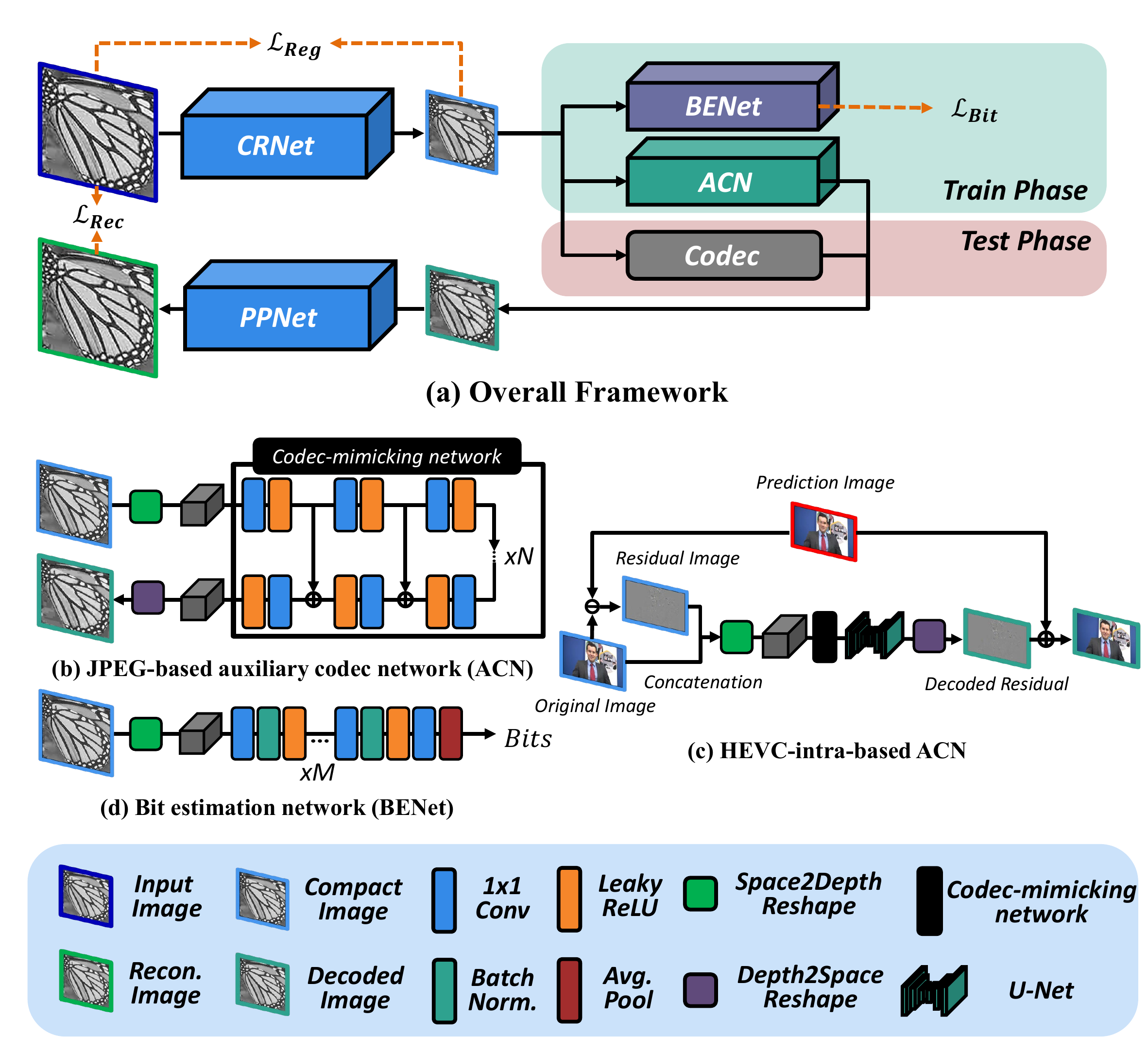}
\caption{(a) Illustration of the end-to-end learning pipeline for image compression. (b)-(d) Detailed structures of the ACN and BENet.}
\label{fig_model}
\end{figure*}

\subsection{Standard Compatible Frameworks based on the Compact Representation Network}
Learning-based image downsampling methods have been less actively researched compared to upsampling methods. These methods can be used as a compact representation of an image without losing important structures and without aliasing effects. Kim \emph{et~al.}~\cite{kim2018task} proposed a task-aware downscaling (TAD) network, and Li \emph{et~al.}~\cite{li2018learning} proposed a compact representation network called CNN-CR to downsample adaptively using joint learning of the downsampling and super-resolution networks. To preserve essential structures in an input image, they both adapted regularization constraints between the input and compact images. Sun \emph{et~al.}~\cite{sun2020learned} proposed a content-adaptive downsampling network using adaptive sampling on the input image to prevent significant changes in the image structure based on a dynamic filter network~\cite{jia2016dynamic}, which does not require regularization loss.

In addition, several approaches can reduce bit rates through downsampling followed by postprocessing in an image compression framework. For example, Afonso \emph{et~al.}~\cite{afonso2018video} and Li \emph{et~al.}~\cite{li2018convolutional} proposed compressing images using the handcrafted downsampling method and restoring images using the deep learning-based super-resolution algorithm. In addition, Jiang \emph{et~al.}~\cite{jiang2017end} proposed an end-to-end framework that consists of three parts: a compact CNN (ComCNN), an image codec, and a reconstruction CNN (RecCNN). The ComCNN produces a compact representation of an input image, and the RecCNN restores the degraded image through compression. The two CNNs are configured with codecs in a compression pipeline to increase coding efficiency.
Jiang~\emph{et~al.} also proposed an iterative optimization algorithm because the end-to-end framework includes a standard codec with nondifferential operations. The CRNet (ComCNN) learns in an end-to-end fashion by connecting directly to the pretrained PPNet (RecCNN). The standard codec is approximated as an identity function in this procedure, which is not optimal in the inference phase.

Unlike \cite{jiang2017end}, Zhao~\emph{et~al.}~\cite{zhao2019learning} proposed a virtual codec neural network (VCNN) to propagate the gradient from the postprocessed image to a preprocessing network called a feature description neural network corresponding to the CRNet in this paper. However, the VCNN and postprocessing neural network corresponding to PPNet are alternately trained because they comprise different pipelines. In addition, the VCNN is designed without careful consideration of the codec structure, making it difficult to guarantee that the correct gradient is propagated through the VCNN. Because the VCNN is used in the training phase and the postprocessing neural network is used when encoding and decoding images in the testing phase, an approximation error of the VCNN causes a train-test discrepancy. In Section~\ref{CRNetbasedmethods}, the CRNet-based compression algorithms will be analyzed and compared in detail.


\section{Proposed Standard Compatible Image Compression Framework}
\subsection{Problem Statement}
We employed a novel image compression framework with the advantages of both the existing standard codec and deep learning-based image processing networks. An important goal in constructing a compression framework was to increase the coding efficiency, which reduces the distortion of the output image and lowers the bit rate of the compressed bitstream. To achieve the optimal parameter of an end-to-end network, we minimized the rate-distortion cost $J = D + \lambda R,$ where $D$ is the distortion, $R$ is the bit rate, and $\lambda$ is the weight of the relative importance between $D$ and $R$. The distortion term $D$ measures how different the reconstructed image is from the original image $\mathbf{x},$ which is defined in the following equation:

\begin{equation}\label{distortion}
D = \delta(\mathbf{x}, g(\Phi(f(\mathbf{x})))),
\end{equation}

\noindent where $f$ denotes the CRNet, $g$ represents the PPNet, and $\Phi$ is the function that generates a reconstruction image from the codec. Moreover, $\delta$ represents a metric that measures the distortion between two images. The bit-rate term $R$ is defined in the following equation:

\begin{equation}\label{birate}
R = \phi(f(\mathbf{x})),
\end{equation}

\noindent where $\phi$ denotes the function that generates the number of bits of the compressed bitstream from the codec. Then, we defined the following objective function to train $f$ and $g$ to minimize the rate-distortion cost:

\begin{equation}\label{idealobjf}
J = \delta(\mathbf{x}, g(\Phi(f(\mathbf{x})))) + \lambda \phi(f(\mathbf{x})).
\end{equation}

Using this objective function, we jointly optimized the CRNet and PPNet to improve the coding efficiency of the codec. However, the codec-related functions $\Phi$ and $\phi$ have a nondifferentiable quantization operation that creates problems for the backpropagation algorithm. The codec-related functions were replaced with a differentiable neural networks: $h$ and $p$, to overcome this problem, as follows:

\begin{equation}\label{objf}
{\theta}_{f}^*, {\theta}_{g}^* \approx \argmin\limits_{{\theta}_{f}, {\theta}_{g}} \delta(\mathbf{x}, g(h(f(\mathbf{x}))) + \lambda p(f(\mathbf{x})),
\end{equation}

\noindent where ${\theta}_{f}$ and ${\theta}_{g}$ are the parameters of the functions $f$ and $g$, respectively. 
In \eqref{objf}, we can reach the ideally optimal solution ${\theta}_{f}$ and ${\theta}_{g}$, if the two neural networks, $h$ and $p$ are perfectly modeled as real codec modules: $\Phi$ and $\phi$. 
All parts of the objective function are composed of learnable neural networks, enabling backpropagation in the end-to-end learning scheme.

We defined $h$ as the ACN and $p$ as BENet in this paper. Both the ACN and BENet have fixed parameters in the process of optimizing \eqref{objf}.
The overall pipeline of the proposed compression framework is illustrated in detail in Fig.~\ref{fig_model} (a). 
The original image passes through the CRNet and is expressed as a compact image to reduce the amount of information. Next, the BENet calculates the number of predicted bits, and the ACN generates an imitated decoded image from the compact image. 
Finally, the PPNet performs restoration of the original image. 
The codec imitation module consisting of the ACN and BENet was used only for training and can be used as a gradient path for training CRNet. In the testing phase, CRNet and PPNet were used with existing codecs, such as conventional preprocessing and postprocessing modules.

\subsection{Auxiliary Codec Network}
The parameter values ${\theta}_{f}^*$ and ${\theta}_{g}^*$ were obtained by optimizing the objective function in \eqref{objf} using the approximation (imitation) function of the codec modules: $\Phi$ and $\phi$. 
However, the obtained parameter may be different when the actual codec is applied. The output of $h$ should be as close as possible to the actual codec module $\Phi$ to reduce these differences and perform optimal learning. In this section, we propose novel CNN architectures of the ACN that closely approximate two typical standard codecs, JPEG and HEVC intra coding (HEVC-intra). The objective function to train the ACNs is as follows:

\begin{equation}\label{objfacn}
{\theta}_{h}^* = \argmin\limits_{{\theta}_{h}} {\lVert h(\mathbf{x}) - \Phi(\mathbf{x}) \rVert}^2_2.
\end{equation}

The architectures of networks imitating JPEG and HEVC-intra are depicted in Fig.~\ref{fig_model} (b) and (c). In the JPEG codec, an original image is divided into $8\times8$ nonoverlapping blocks, and it independently processes each block. Referring to the characteristics of this JPEG compression, an input image divided into fixed $8\times8$ blocks was calculated to change the axis through the Space2Depth and Depth2Space operations proposed in \cite{shi2016real}. The Space2Depth operation performs data conversion from the blockwise spatial axis to the channel axis. For the JPEG-based ACN, pixel data in an $8\times8$ block were rearranged and converted to $64$ channels. For example, if the input image with $C$ color channels, height $H$, and width $W$ is represented as a data tensor of $H\times W\times C$, it is reshaped into a data tensor of $\frac{H}{8}\times\frac{W}{8}\times64C$ through the Space2Depth operation. Next, the pixel in each block independently forms a full connection while passing through $1\times1$ convolutional layers and skip connections, preventing a large loss of information as the input image passes through the deep network.

In the HEVC intra-encoding process, more complex encoding is performed using intra-prediction methods, and the size of the coding block is variable. Because the prediction image for reducing the redundancy of a block is not generated inside the block itself but from other blocks, it is difficult to approximate a prediction-based codec only with the previously proposed block-based Space2Depth operation, which performs a connection only within blocks.

Therefore, a more complex structure for the HEVC-intra-based ACN is proposed, as illustrated in Fig.~\ref{fig_model} (c). An original and prediction image were concatenated along the channel axis and provided as input. The prediction image was generated from the real codec and makes the HEVC-intra-based ACN focus only on generating the coded prediction residual. Continuously generating the prediction image using the real codec in the end-to-end training process does not affect the actual encoding and decoding processes because the ACN is not used at the inference phase.

For the HEVC-intra-based ACN, we also divided the block size of $8\times8$ with the same size as JPEG-based ACN because the default value of the minimum CU unit was set to $8$ in the default profile of the HEVC reference software HM $16.20$ \cite{mccann2014high, HMreference2019}. However, the HEVC performs transform and quantization with variable block sizes, so the structure of the HEVC-intra-based ACN comprises the combination of a JPEG-based ACN and additional CNNs based on the U-Net structure~\cite{ronneberger2015u}. The proposed structure of the HEVC-intra-based ACN with additional CNNs overcomes the limitations of the JPEG-based ACN, in which the weights were connected only in blocks of a fixed size by enabling a connection between divided blocks. Finally, an imitated decoded image of the HEVC-intra-based ACN was obtained from the sum of the residual image generated by the combined CNNs and the input prediction image. 

\subsection{Bit Estimation Network}
As all parts of the objective function should comprise differentiable functions, we approximated the bit-rate term expressed as the size of a bitstream generated from the existing codec encoder $\phi$ using the deep learning network function $p$. The objective function to train the BENet is as follows:

\begin{equation}\label{objfbe}
{\theta}_{p}^* = \argmin\limits_{{\theta}_{p}} {\lVert p(\mathbf{x}) - \phi(\mathbf{x}) \rVert}^2_2.
\end{equation}

We propose the architecture of the BENet, which predicts the number of bits for end-to-end learning, considering the bit-rate term in \eqref{objf}. The structure of the BENet is presented in Fig.~\ref{fig_model} (d). The depth of the BENet $M$ was set to $10$ with a sufficiently small error to proceed with end-to-end learning based on the results of the ablation study in Section~\ref{AblationStudy}. An input image of the BENet was rearranged on the channel axis in the same way as the ACN and generates a predicted value through the $1\times1$ convolution and global average pooling. The BENet can predict the size of the bitstream generated after encoding from the input image. The CNN structure can backpropagate the gradient from the bit rate described in the objective function. With the BENet comprising differentiable operations, gradients of the network weights can be calculated from approximate bits. Finally, a gradient-based optimization algorithm can be applied to optimize the objective function \eqref{objf}.


\section{Network Optimization}
The end-to-end network was optimized using approximate functions, as in \eqref{objf}. An error may occur between weights obtained through end-to-end learning and ideal optimized weights when using this function. In this section, to reduce this error and more closely reach the ideal parameters of the CRNet and PPNet, we propose an appropriate loss function and effective training strategy, including a pretraining method for each network and an iterative fine-tuning update algorithm.

\subsection{Loss Function}
The goal of end-to-end network training is to improve the coding efficiency; that is, the reconstructed image through the PPNet should be closer to the original image $\mathbf{x}$, and the number of bits generated by the standard encoder $\phi(\mathbf{x})$ should be reduced. The distortion is defined as the mean squared error of the original image $\mathbf{x}$ and reconstructed image $\hat{\mathbf{x}}$. The result obtained using the distortion calculation is defined as the reconstruction loss ${\mathcal{L}}_{rec}$ as presented in the following equation:

\begin{equation}\label{Recloss}
{\mathcal{L}}_{rec} = {\lVert \mathbf{x} - \hat{\mathbf{x}} \rVert}^2_2.
\end{equation}

Next, \eqref{Bitloss} defines the bit loss ${\mathcal{L}}_{bit}$ to reduce the number of bits:

\begin{equation}\label{Bitloss}
{\mathcal{L}}_{bit} = p(f(\mathbf{x})).
\end{equation}

The end-to-end deep learning model does not analyze the characteristics of the image with human intuition but only updates the weight in a direction that reduces the objective function. Therefore, it is difficult to determine the pattern of intermediate products of each module in a pipeline. As the parameter of $f$ is updated as the training progresses, the approximation performance of $h$ and $p$ gradually decreases because the unseen compact image is input into the pretrained $h$ and $p$. The regularization loss proposed in \cite{li2018learning}, which is called guide loss in \cite{kim2018task}, solves the degradation of the approximation accuracy as follows:

\begin{equation}\label{Regloss}
{\mathcal{L}}_{reg} = {\lVert f(\mathbf{x}) - F_{s}(\mathbf{x}) \rVert}^2_2,
\end{equation}

\noindent where $F$ denotes the bicubic downsampling function with scale factor $s$. Regularization loss ${\mathcal{L}}_{reg}$ helps the compact image generated by $f$ preserve the statistical characteristics of natural images and the low-frequency information of the original image. The total loss combined with the above three loss types is expressed in the following equation:

\begin{equation}\label{Totalloss}
\mathcal{L}_{total} = {\mathcal{L}}_{rec} + \lambda_{bit}{\mathcal{L}}_{bit} + \lambda_{reg}{\mathcal{L}}_{reg},
\end{equation}

\noindent where $\lambda_{bit}$ and $\lambda_{reg}$ are trade-off weights for balancing each loss term. The compression efficiency results for $\lambda_{bit}$ and $\lambda_{reg}$ are discussed in more detail in Section~\ref{Exp:Loss}.

\subsection{Training Strategy}\label{TrainingStrategy}
\subsubsection{Pretraining}\label{pretraining}
The approximation networks, ACN and BENet, are auxiliary networks used only for end-to-end learning and perform only gradient backpropagation. These networks are already given the role of codec imitation and bit prediction, regardless of the learning direction of end-to-end learning. Therefore, the two networks should be pretrained. The ACN can be pretrained using \eqref{objfacn}, and the BENet can be pretrained using \eqref{objfbe}.

The CRNet processes the input image before the standard codec, and the PPNet processes the decoded image generated by the standard codec. The proposed compression framework aims to increase the coding efficiency by lowering the resolution of the input and output images. Because it is already known that the CRNet and PPNet should be able to change the resolution of an image and remove compression artifacts, it is beneficial to define the initial state of the CRNet and PPNet through pretraining. The initial state is defined in the following equations:

\begin{equation}\label{InitialstateF}
{\theta}_{f}^0 = \argmin\limits_{{\theta}_{f}} {\lVert f(\mathbf{x}) - F_{s}(\mathbf{x}) \rVert}^2_2,
\end{equation}

\begin{equation}\label{InitialstateG}
{\theta}_{g}^0 = \argmin\limits_{{\theta}_{g}} {\lVert g(\Phi(F_{s}(\mathbf{x}))) - \mathbf{x} \rVert}^2_2.
\end{equation}

We trained the CRNet to output a bicubic downsampling image at the initial state and trained the PPNet to restore a degraded bicubic downsampled image to the original image. The pretraining strategy for all networks provides a good initialization point, making the optimal parameter closer to the ideal and obtaining a faster convergence rate. This result is verified using a comparative experiment in Section~\ref{Exp:Pretraining}.

\begin{algorithm}
\SetAlgoLined
\KwInput{Original image: $\mathbf{x}$; Batch size: $K$;}
Pretrain $h$ and $p$ by optimizing Eqs. \eqref{objfacn} and \eqref{objfbe}\
Initialize ${\theta}_{f}^0$ and ${\theta}_{g}^0$ by optimizing Eqs. \eqref{InitialstateF} and \eqref{InitialstateG}\
  \For{each minibatch iteration}{
  Update ${\theta}_{f}$ and ${\theta}_{g}$ via joint learning with fixed $h$ and $p$, referring to the objective function \eqref{Totalloss}\
  
  \For{$i = 1 \gets K$}{
   Generate the training data group containing a compact image $f(\mathbf{x})$, a decoded image $\Phi(f(\mathbf{x}))$, and the number of bits $\phi(f(\mathbf{x}))$ from the codec\
   }
  Update the parameters of $h$ and $p$ using grouped data $\{f(\mathbf{x}), \Phi(f(\mathbf{x})), \phi(f(\mathbf{x}))\}$ using Eqs. \eqref{objfacn} and \eqref{objfbe}\
 }
\KwReturn{${\theta}_{f}$, ${\theta}_{g}$}
 \caption{Training algorithm of the proposed compression framework}
 \label{alg:iterative}
\end{algorithm}

\begin{figure}
\centering
   \subfloat[Alternate learning with identity function approximation~\cite{jiang2017end}]{\includegraphics[width=0.99\linewidth]{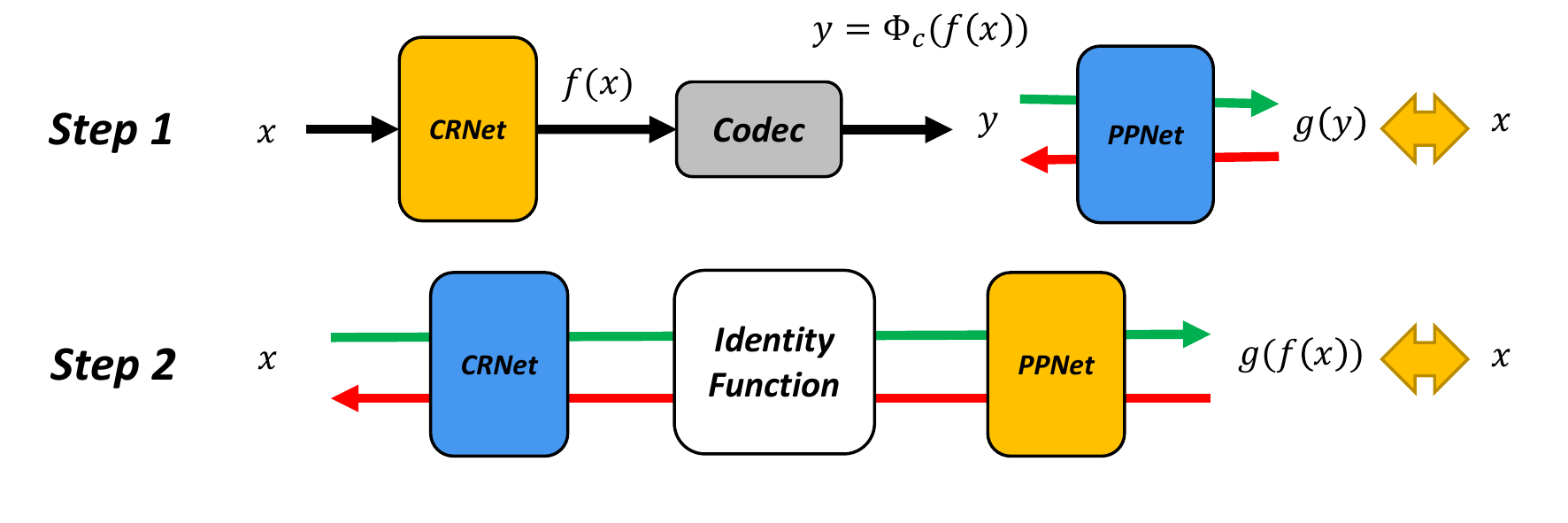}}\
   \subfloat[Alternate learning with a virtual codec neural network (VCNN)~\cite{zhao2019learning}]{\includegraphics[width=0.99\linewidth]{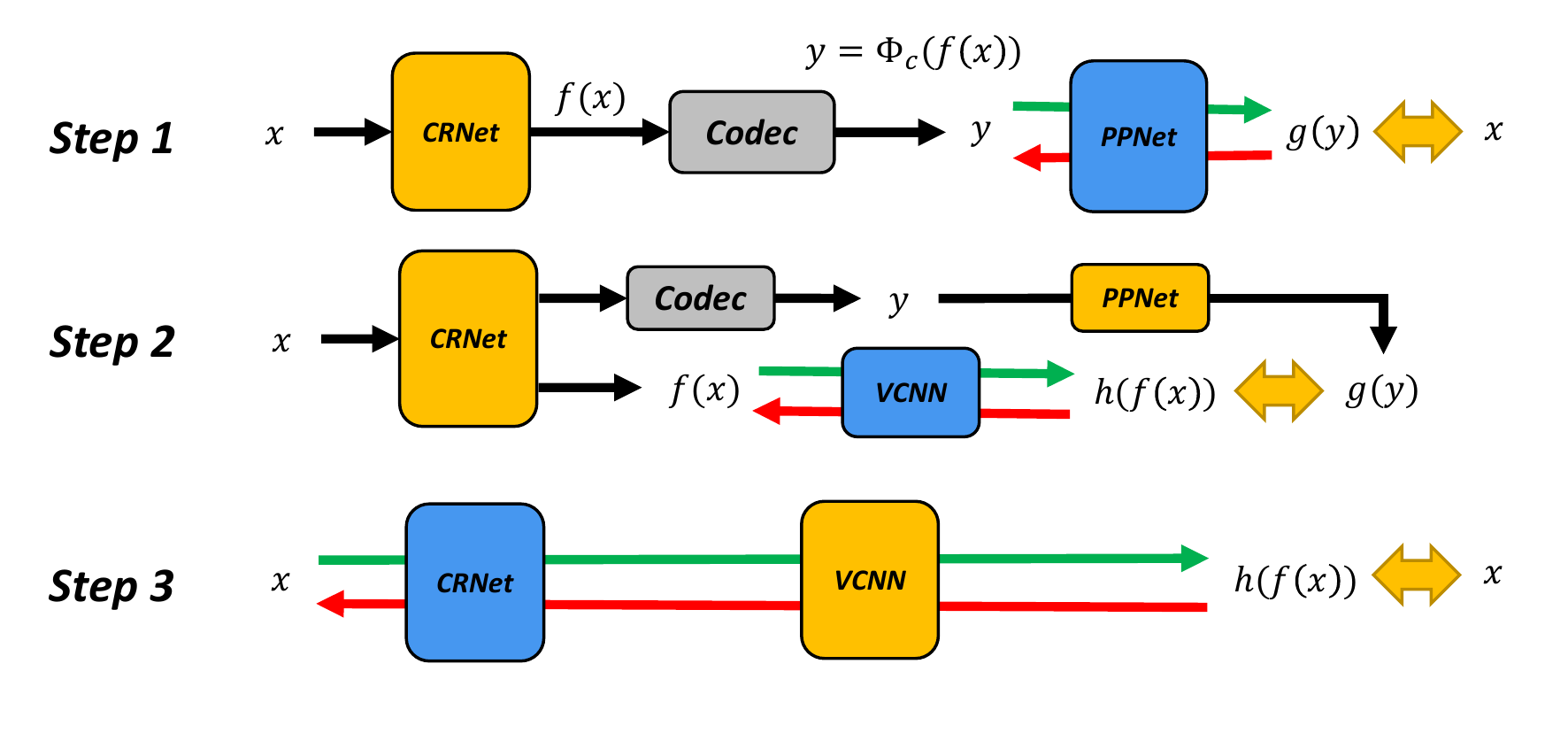}}\
   \subfloat[Simultaneous learning with the proposed auxiliary codec network (ACN)]{\includegraphics[width=0.99\linewidth]{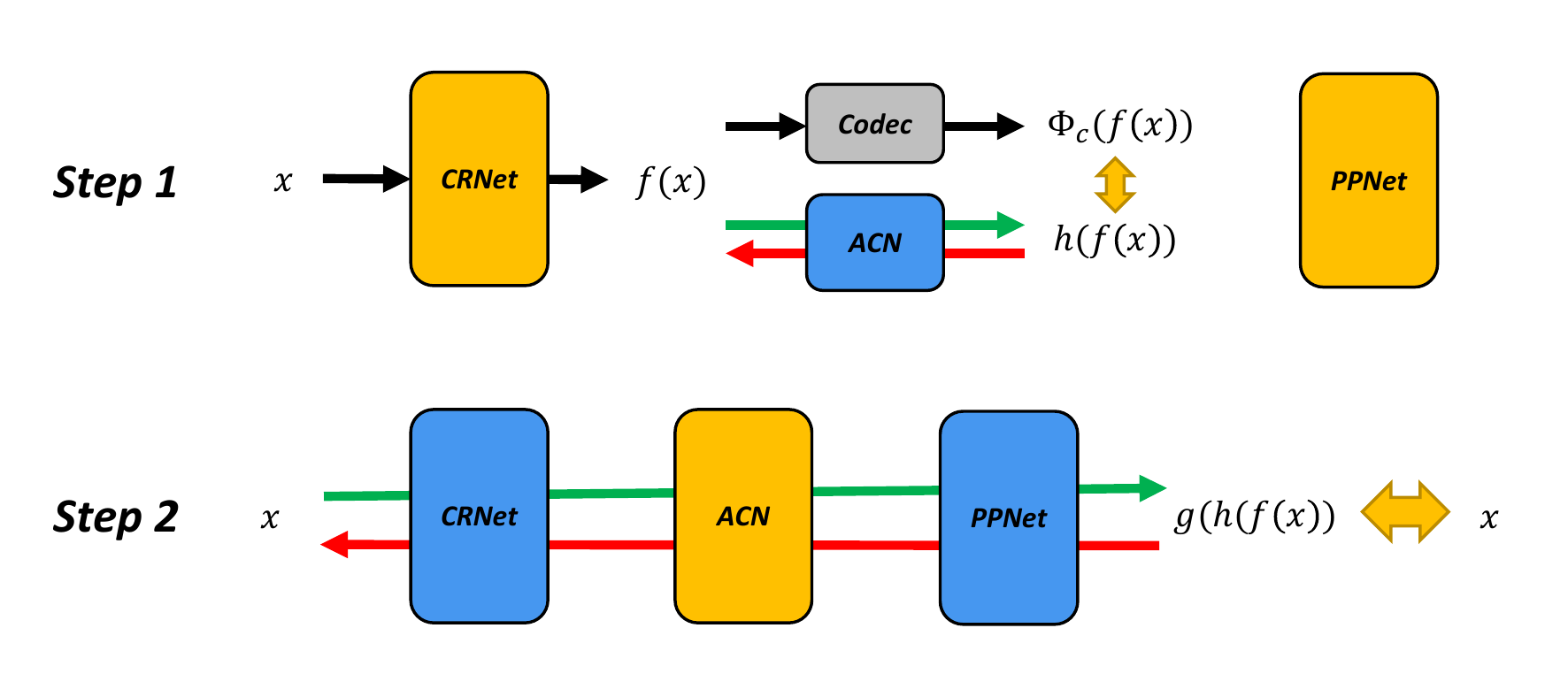}}
   \caption{Comparison of the learning process of compact representation network (CRNet)-based methods. The blue module indicates the status updated in each step, and the yellow module indicates the fixed status. Green and red arrows indicate a forward pass (or inference) and a backward pass (or gradients) to train the module, respectively. The yellow double arrow indicates the argument of the loss function for the backward pass.}
   \label{fig:comparisonofSOTA}
\end{figure}

\subsubsection{Iterative Fine-tuning Updating}\label{iteratvielearning}
As the fine-tuning of the end-to-end model progresses, the CRNet learns in the direction of optimizing the objective function. However, as mentioned, the use of the approximation function affects the learning of the entire model. The ACN and BENet, which have fixed weights, gradually decrease the approximation accuracy to the standard codec because the unseen compact image from the CRNet is input as the whole model is trained. Therefore, the ACN and BENet are updated through an iterative learning process, as presented in Algorithm \ref{alg:iterative}. The parameters of the CRNet and PPNet are updated for each cycle of the minibatch, and the ACN and BENet are updated using the output value of the CRNet. By correcting the approximation errors with the proposed algorithm, the parameters of the CRNet and PPNet can be learned more closely to the ideal optimum values ${\theta}_{f}^*$ and ${\theta}_{g}^*$.

\subsubsection{Comparison with the other CRNet-based Methods}\label{CRNetbasedmethods}
Recent CRNet-based papers~\cite{jiang2017end, zhao2019learning} proposed a learning strategy that bypasses the nondifferentiable codec, adopting an alternate learning method for CRNet and PPNet. In Fig.~\ref{fig:comparisonofSOTA} (a), which illustrates the learning method of \cite{jiang2017end}, CRNet was learned by directly connecting the PPNet, assuming the codec to be an identity function. On the other hand, in Fig.~\ref{fig:comparisonofSOTA} (b), which illustrates the learning method of \cite{zhao2019learning}, CRNet was trained using the VCNN which generates the image passed through the codec and PPNet. In both methods, CRNet and PPNet are trained alternately, which is an incomplete optimization process.

In this paper, we adopted a method of simultaneously learning an end-to-end network so that the CRNet and PPNet obtain the weights closer to the ideal optimal values. We performed a better optimization learning method for CRNet and PPNet to be optimized simultaneously rather than other alternate learning methods. As illustrated in Fig.~\ref{fig:comparisonofSOTA} (c), CRNet and PPNet are updated simultaneously with a fixed ACN. The comparative experiments for alternate and simultaneous learning are discussed in more detail in Section~\ref{exp:Learningmethod}

\begin{figure}
   \centering
   \subfloat[][]{\includegraphics[width=.5\linewidth]{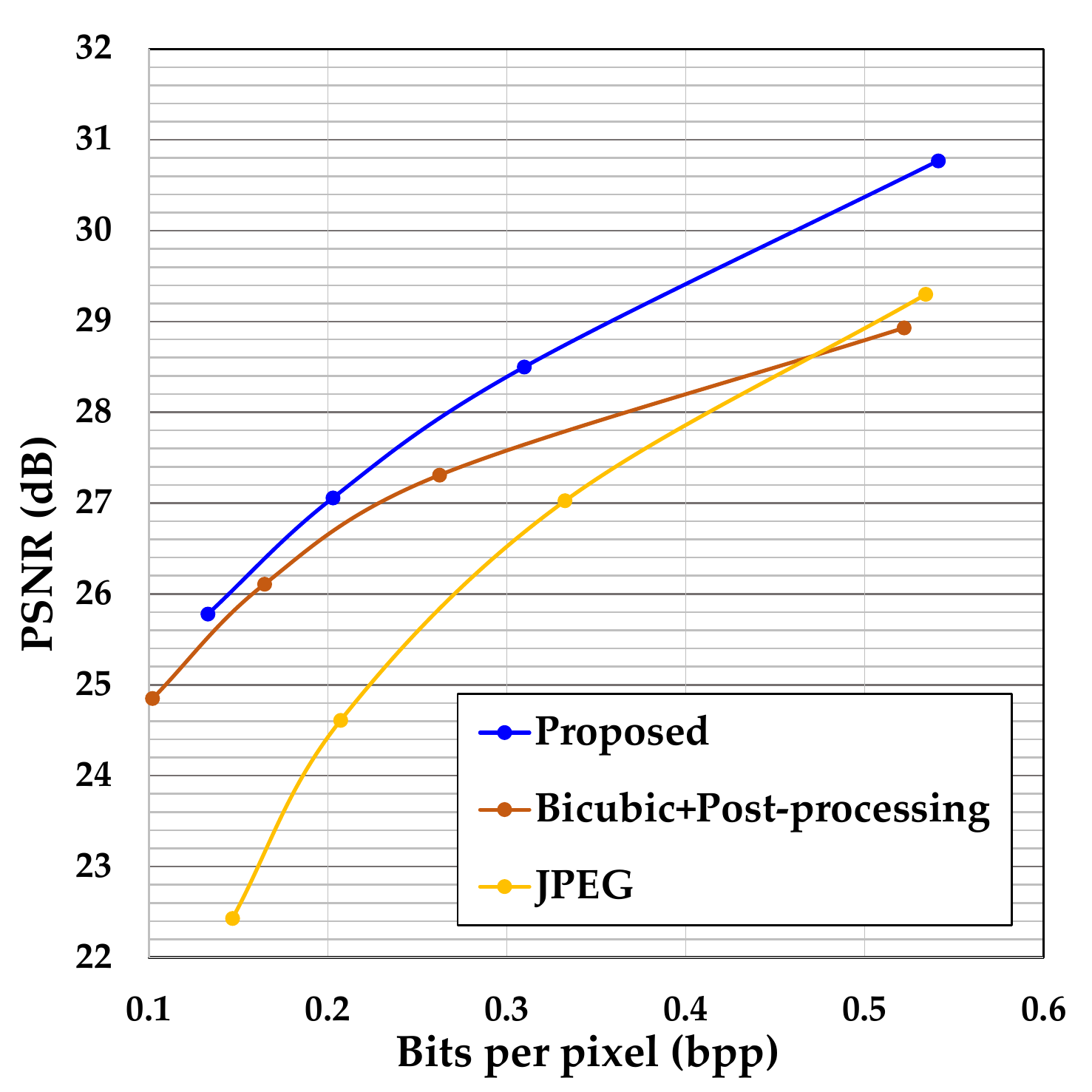}}
   \subfloat[][]{\includegraphics[width=.5\linewidth]{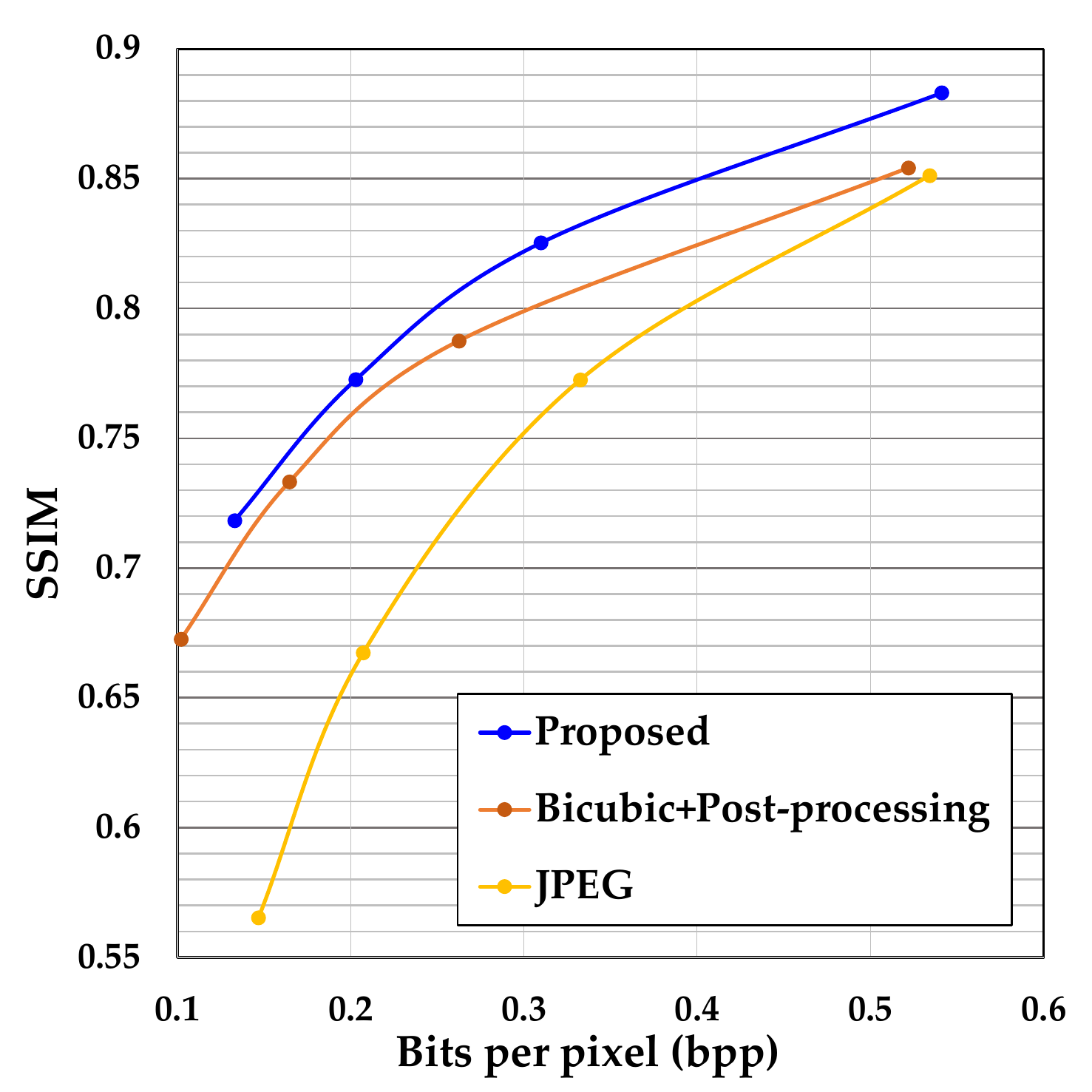}} \\
   \subfloat[][]{\includegraphics[width=.5\linewidth]{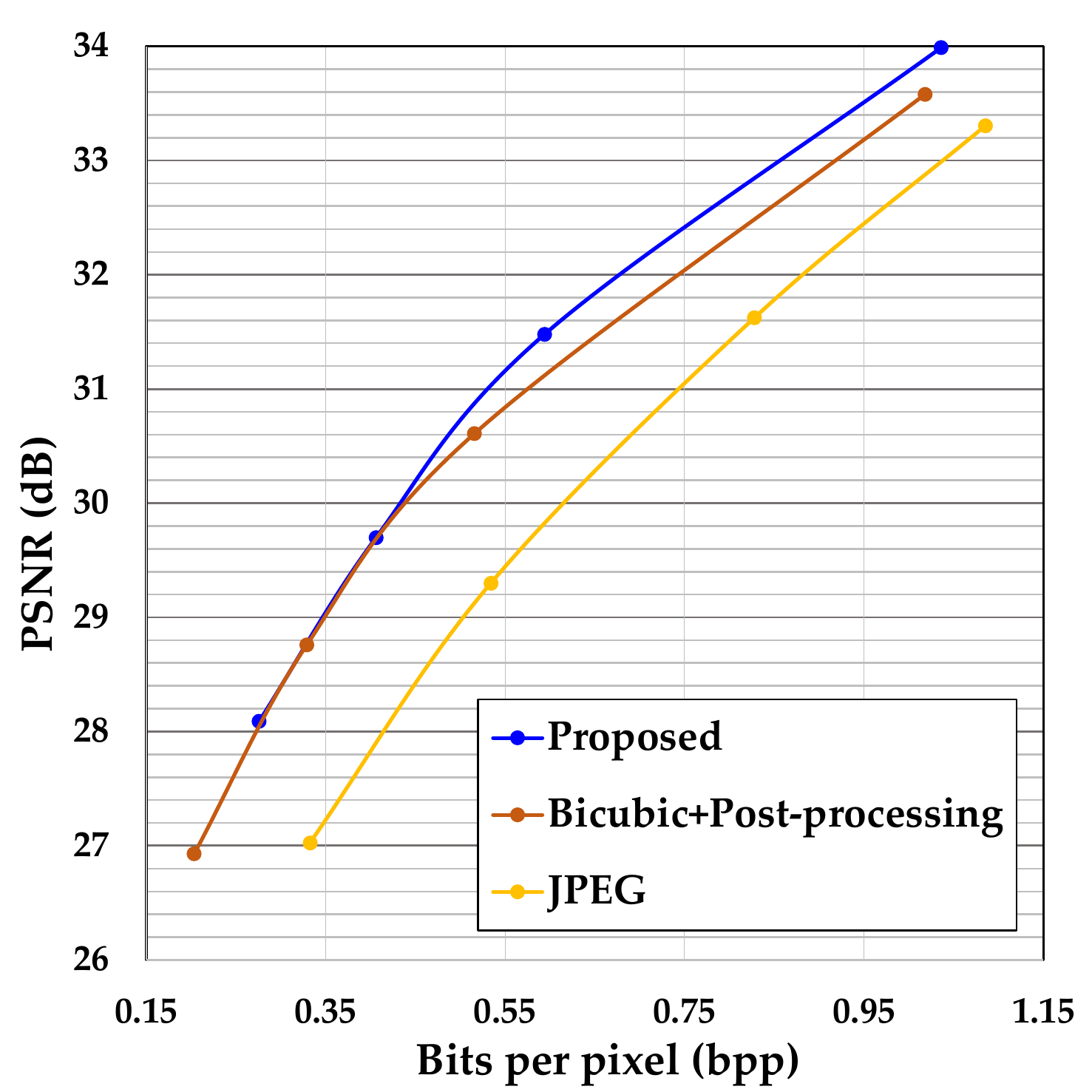}} 
   \subfloat[][]{\includegraphics[width=.5\linewidth]{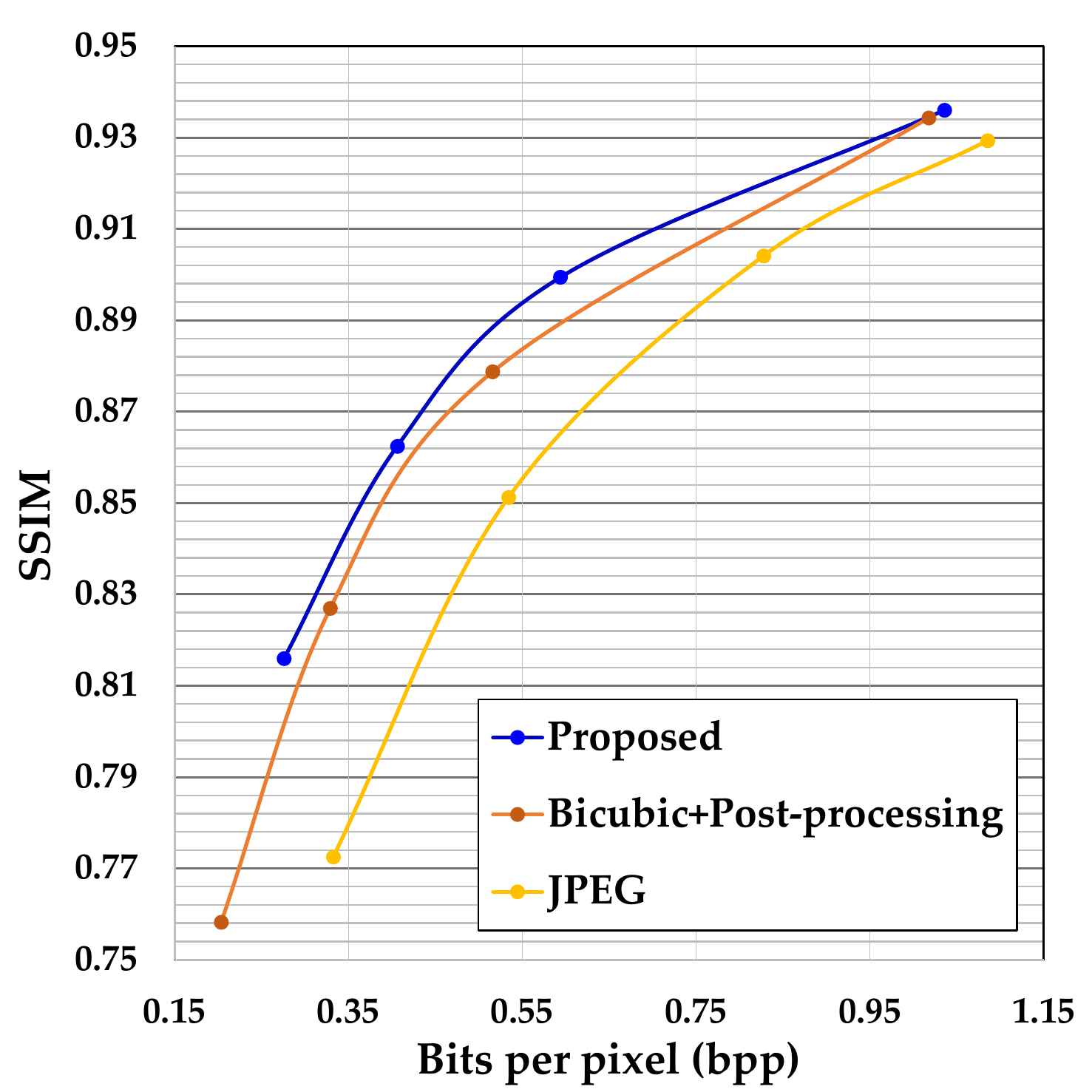}} \\
   \subfloat[][]{\includegraphics[width=.5\linewidth]{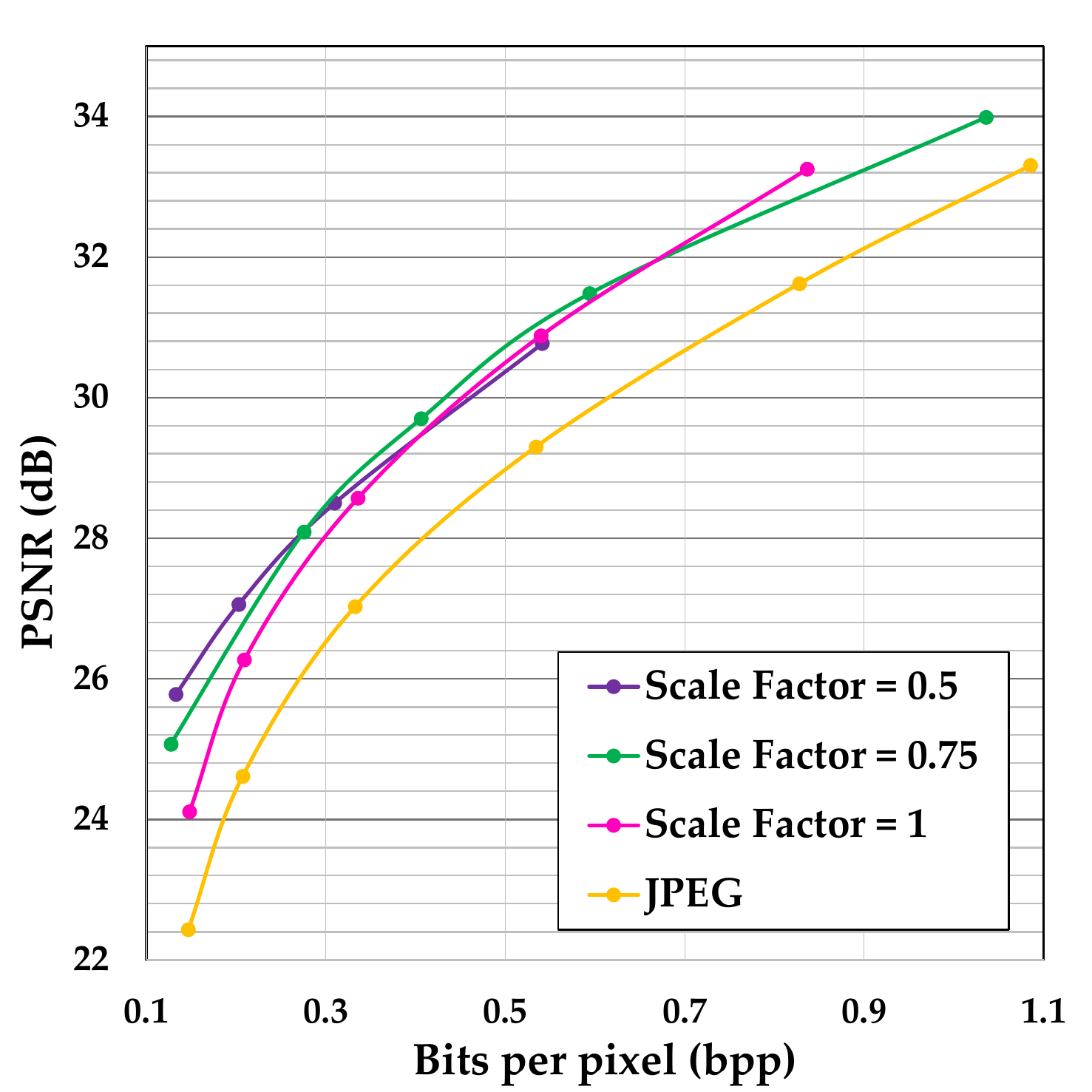}}
   \subfloat[][]{\includegraphics[width=.5\linewidth]{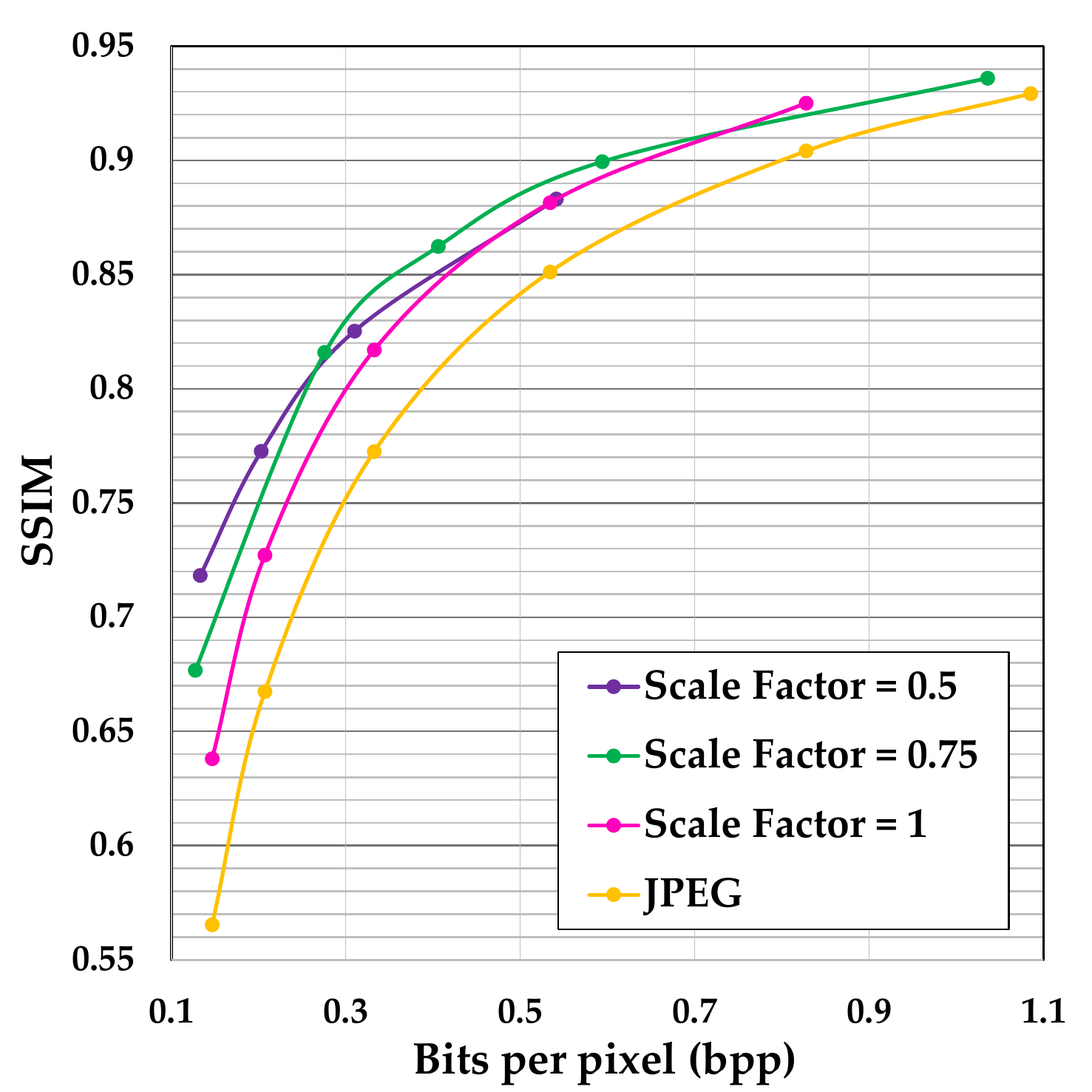}} 
   \caption{Rate-distortion curve of the proposed CRNet and bicubic downsampling in two downsampling scale factors and comparison according to a scale factor. (a),(b) Scale factor$= 0.5$; (c),(d) Scale factor$= 0.75$; and (e),(f) all scale factors.}
   \label{fig:abl_CRnetcomp}
\end{figure}

\begin{figure*}
   \centering
   \subfloat[][]{\includegraphics[width=0.99\linewidth]{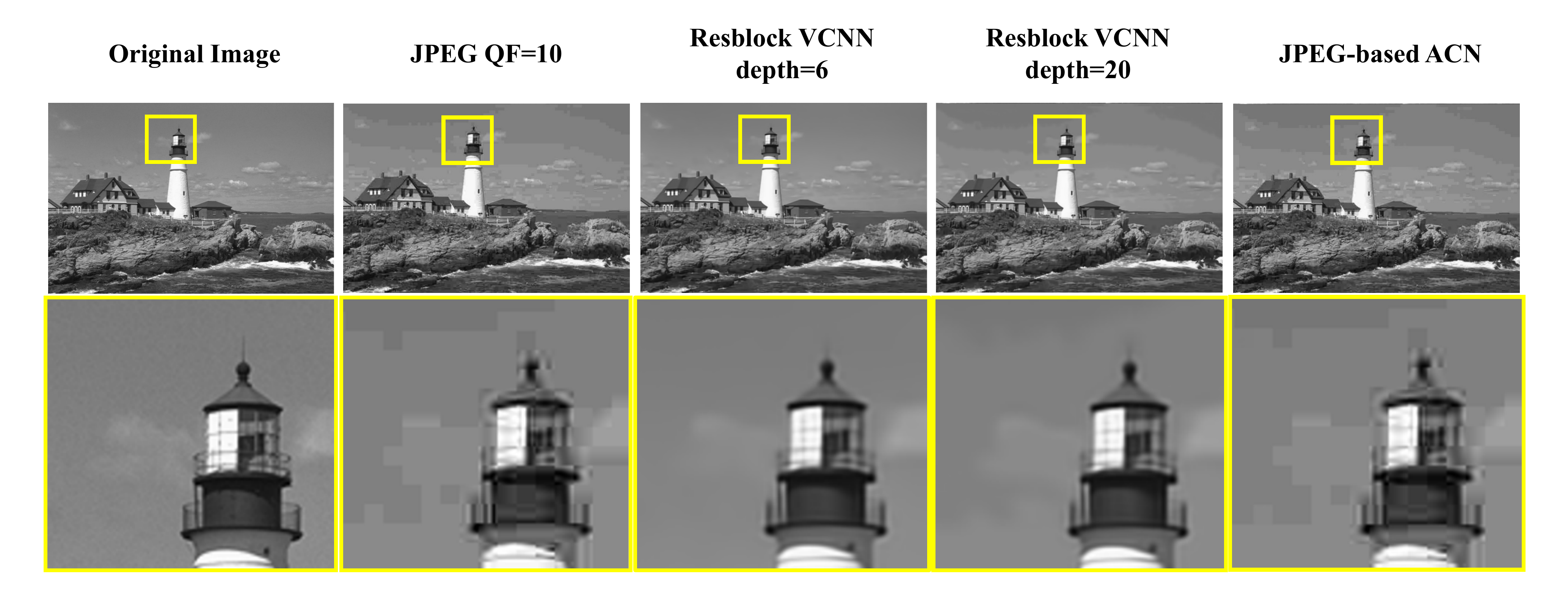}}\
   \subfloat[][]{\includegraphics[width=0.99\linewidth]{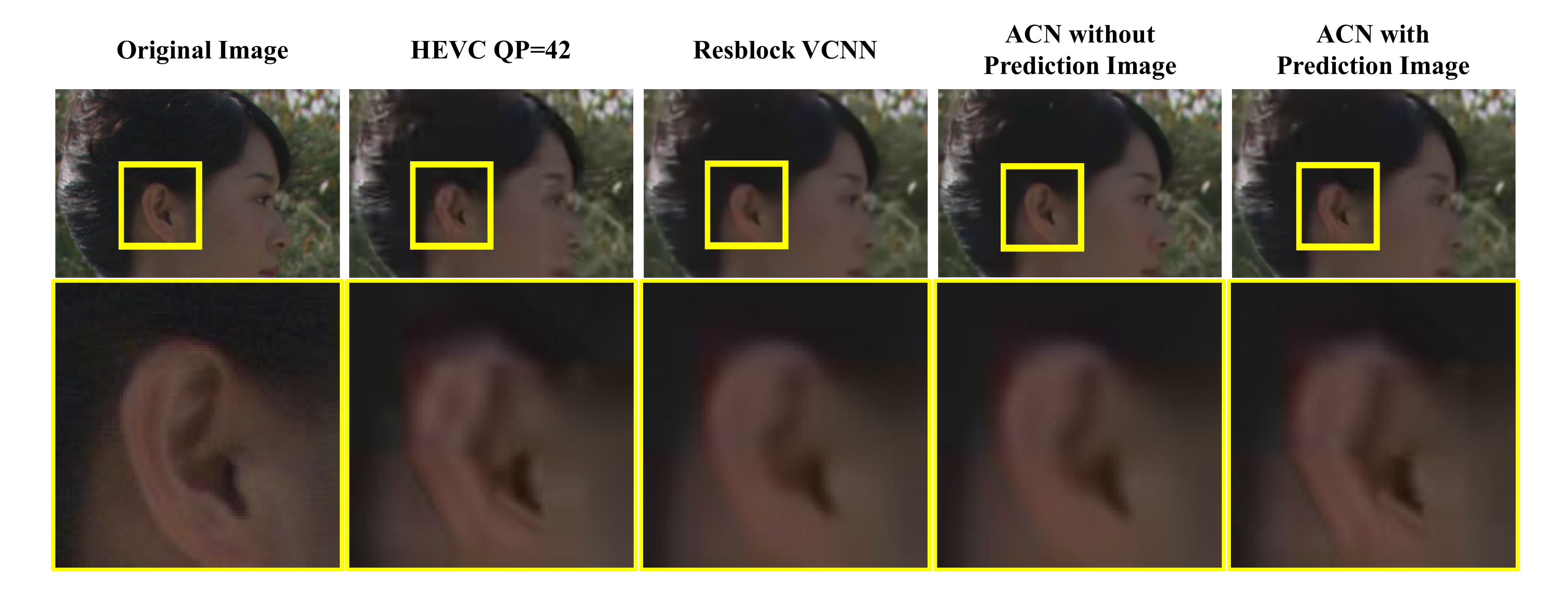}}
   
   \caption{Visual comparison of different codec-mimicking network structures on the (a) \textit{Lighthouse} image of the LIVE1 dataset at a quality factor of $10$, and (b) \textit{Kimono} of the HEVC Test Sequence~\cite{bossen2013common} at an HEVC quality parameter of $42$. The result of JPEG-based ACN has blocking artifacts and ringing artifacts around edges similar to the image compressed with JPEG. The result of ACN with prediction image, unlike other structures, is less blurry and has compression artifacts similar to HEVC.}
   \label{fig:abl_ACNmodelcomp}
\end{figure*}

\begin{table*}
  \centering
  \caption{Quantitative Peak Signal-to-Noise Ratio (PSNR; dB) Comparison of JPEG Imitation Performance by Different Codec-mimicking network Structures on Set14 and LIVE1 Datasets}
    \begin{tabular}{ccccccccc}
    \toprule
    \toprule
    \multirow{2}[4]{*}{vs JPEG} & \multicolumn{2}{c}{$QF=10$} & \multicolumn{2}{c}{$QF=20$} & \multicolumn{2}{c}{$QF=40$} & \multicolumn{2}{c}{$QF=80$} \\
\cmidrule{2-9}      & Set14 & LIVE1 & Set14 & LIVE1 & Set14 & LIVE1 & Set14 & LIVE1 \\
    \midrule
    Original image & 27.49  & 27.03  & 29.85  & 29.30  & 32.20  & 31.62  & 37.00  & 36.32  \\
    VCNN~\cite{zhao2019learning} $depth=6$ & 29.89  & 29.56  & 32.15  & 31.60  & 34.29  & 33.60  & 37.98  & 37.49  \\
    VCNN~\cite{zhao2019learning} $depth=20$ & 29.93  & 29.68  & 32.33  & 31.77  & 34.37  & 33.67  & 37.98  & 37.49  \\
    JPEG-based ACN $depth\,N=9$ & 43.85  & 44.50  & 39.82  & 39.82  & 40.34  & 40.29  & 39.53  & 38.97  \\
    JPEG-based ACN $depth\,N=11$ & 45.24  & 45.41  & 41.78  & 41.75  & 40.33  & 40.25  & 39.76  & 39.25  \\
    JPEG-based ACN $depth\,N=12$ & \textbf{46.07}  & \textbf{46.27}  & \textbf{42.32}  & \textbf{42.28}  & \textbf{41.27}  & \textbf{41.25}  & \textbf{43.08}  & \textbf{42.61}  \\

    \bottomrule
    \bottomrule
    \end{tabular}%
  \label{tab:ACNmodel}%
\end{table*}%

\begin{table*}
  \centering
  \caption{Quantitative Peak Signal-to-Noise Ratio (PSNR; dB) Comparison of HEVC Imitation Performance by Different Codec-mimicking network Structures on HEVC Test Sequences~\cite{bossen2013common}}
    \begin{tabular}{ccccc}
    \toprule
    \toprule
    vs HEVC & $QP=32$ & $QP=37$ & $QP=42$ & $QP=47$ \\
    \midrule
    Original image & 36.07 & 33.06 & 30.13 & 27.40 \\
    VCNN~\cite{zhao2019learning} & 38.54 & 36.17 & 33.94 & 32.07 \\
    ACN without prediction image & 39.12 & 36.63 & 34.37 & 32.36 \\
    ACN with prediction image & \textbf{40.87} & \textbf{39.09} & \textbf{37.53} & \textbf{36.06} \\
    \bottomrule
    \bottomrule
    \end{tabular}%
  \label{tab:HEVCACN}%
\end{table*}%

\section{Experimental Results}

\subsection{Setting}\label{Setting}
Previous studies based on the CRNet and PPNet \cite{jiang2017end, zhao2019learning} have displayed performance improvement in various image codecs, such as JPEG, JPEG2000 \cite{rabbani2002jpeg2000} and BPG~\cite {bellard2015bpg}. In this paper, we experimented based on the traditional and widely used JPEG image codec to demonstrate the effectiveness of the proposed method. Furthermore, the HEVC standard based on HEVC reference software HM $16.20$ \cite{mccann2014high, HMreference2019} is additionally used for codec modeling to demonstrate the possibility of the extension to general codecs.

The architecture of the CRNet adopts the TAD structure~\cite{kim2018task} and the PPNet adopts an enhanced deep super-resolution network (EDSR)~\cite{lim2017enhanced} as the baseline structure. Because the CRNet performs downsampling and the PPNet performs upsampling, the TAD and EDSR are representative convolutional networks for resizing the input image.

The DIV2K dataset \cite{timofte2017ntire} was used to train the neural network of the proposed framework. The DIV2K training set consists of 800 high-resolution images. Additionally, we used 123,403 images from the COCO 2017 dataset \cite{lin2014microsoft} to pretrain the ACN and BENet to learn more diverse patterns and increase the approximation accuracy of the codec imitation module. We divided the original images into $128\times128$ patches for training. We used the Adam optimizer \cite{kingma2015adam} with $\beta_1 = 0.9$ and $\beta_2 = 0.99$ to train the networks. We pretrained each network module with the size of the minibatch set to $16$ and the learning rate to $1\times10^{-4}$. After pretraining, the learning rate was set to $5\times10^{-5}$, and for the ACN and BENet update, the learning rate was set to $5\times10^{-6}$.

Two parameters of the loss function, $\lambda_{bit}$ and $\lambda_{reg}$, were empirically determined. We trained the model per JPEG quality factor (QF) from $10$ to $80$. A higher QF indicates a higher bit rate and a larger value of $\lambda_{bit}$ results good performance in a high bit-rate environment. Specifically, $\lambda_{bit}$ is set to $2\times10^{-4}$ when the QF is $10$, $1\times10^{-4}$ when the QF is $20$, and $3\times10^{-5}$ when the QF is 40 and 80. The details regarding setting the weights are described in the experimental results of the ablation study.
For JPEG-based model testing, two benchmark datasets, Set14 \cite{zeyde2010single} and LIVE1 \cite{sheikh2006statistical} were used. 

In the HEVC-intra-based model, applying adaptive weights according to the quantization parameter (QP) does not greatly facilitate performance improvement; thus, $\lambda_{bit}$ and $\lambda_{reg}$ are set to $5\times10^{-5}$ and $1$, respectively, for all QPs.
The HEVC-intra-based model is implemented in the HEVC reference software HM $16.20$ \cite{mccann2014high, HMreference2019} with the all intra configuration and tested based on test conditions, configurations, and sequences proposed by the Joint Collaborative Team on Video Coding~\cite{bossen2013common}. The test sequences can be divided into Classes A, B, C, D, and E according to the spatial resolution. When evaluating the compression performance of HEVC-intra-based models, the results are expressed in terms of the Bjøntegaard delta (BD) rate~\cite{bjontegaard2001calculation} reductions for the luma component. In both codec model test situations, we adopted the peak signal-to-noise ratio (PSNR) and the structural similarity index measure (SSIM)~\cite{wang2004image} as image quality evaluation metrics. 

\begin{figure}
   \centering
   \subfloat[][]{\includegraphics[width=0.99\linewidth]{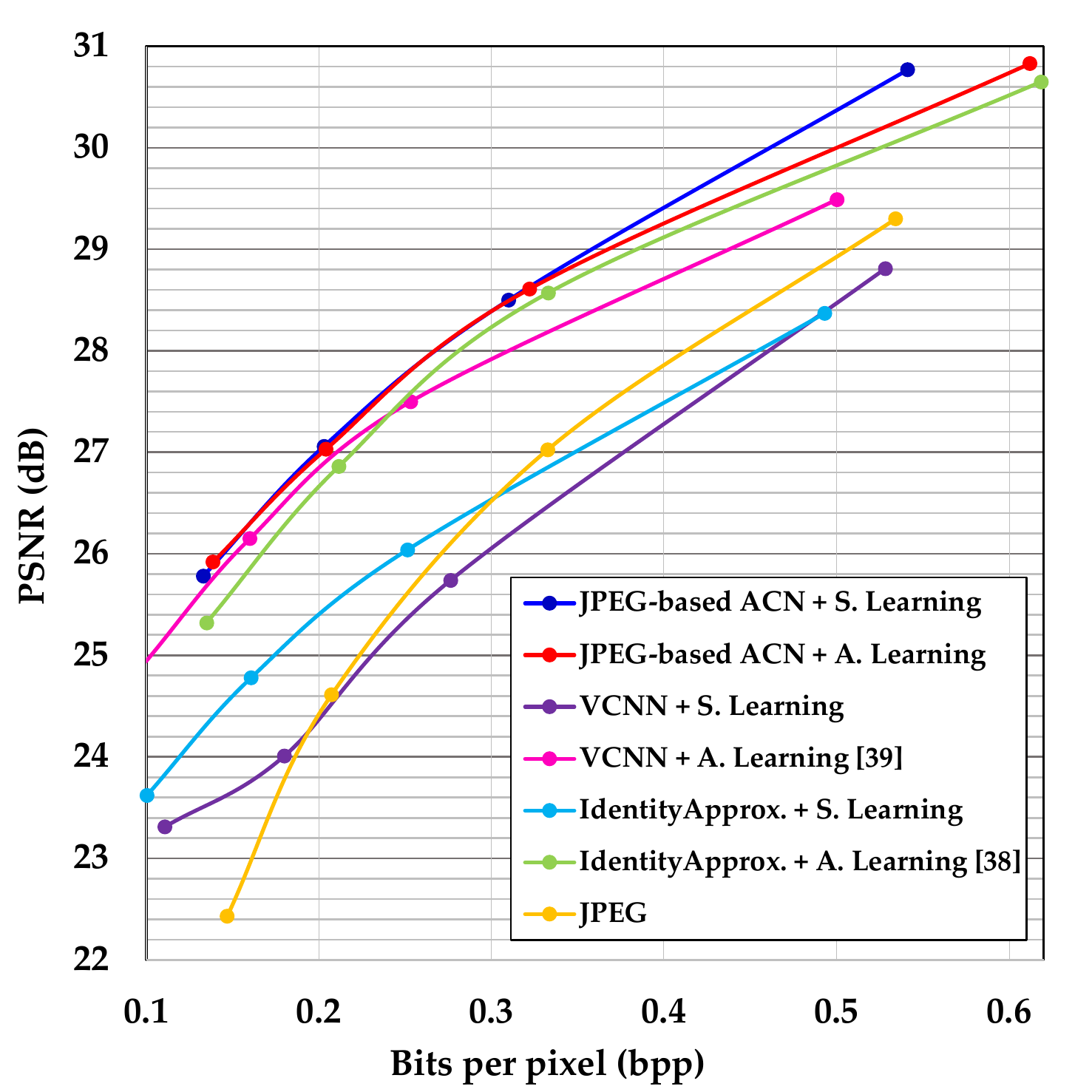}}\
   \subfloat[][]{\includegraphics[width=0.99\linewidth]{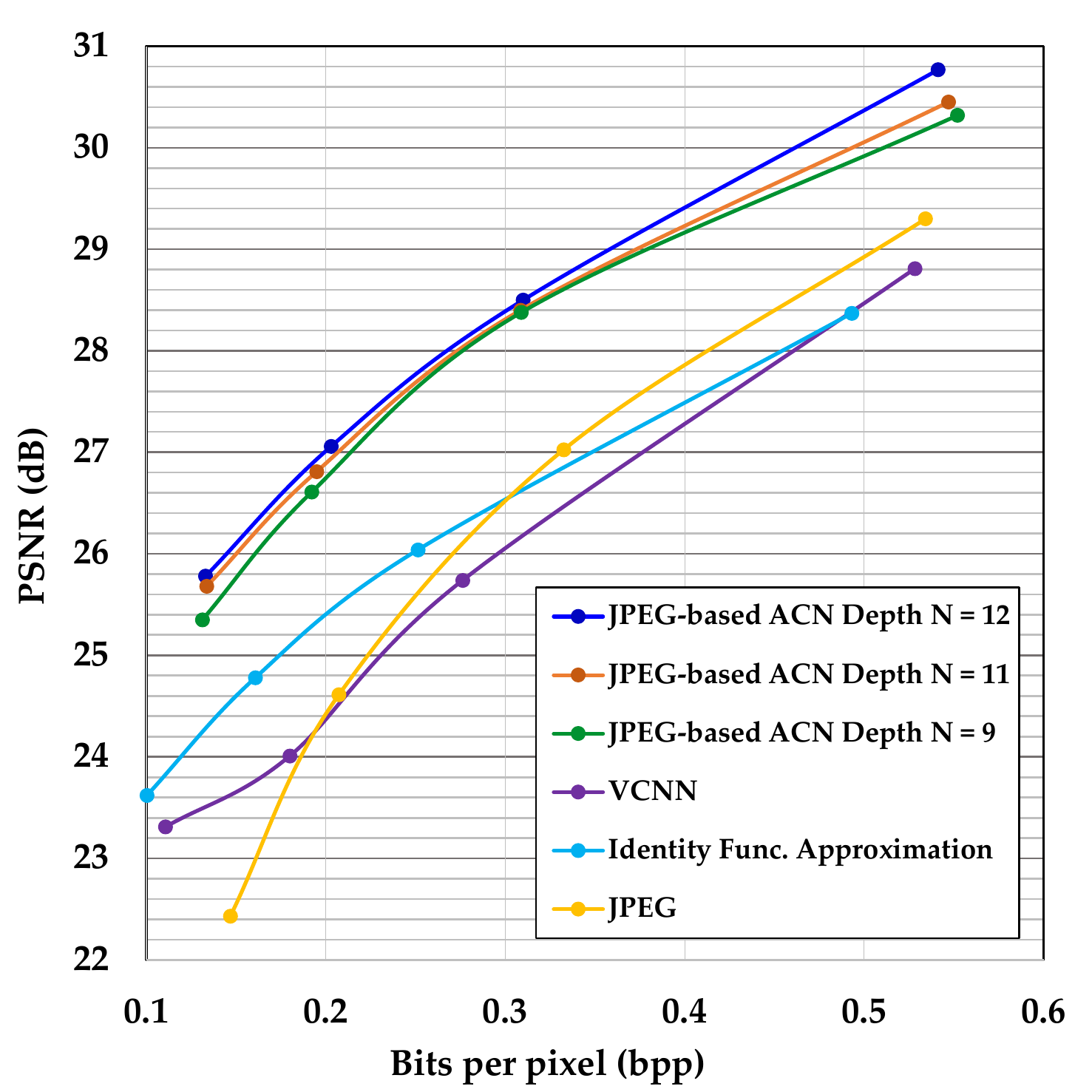}}
   \caption{Rate-distortion curve comparison of compression methods using preprocessing and postprocessing networks. (a) Comparison results according to network architectures and learning methods (simultaneous learning (S. Learning) or alternating learning (A. Learning)), (b) Comparison result of the codec-mimicking networks with simultaneous learning.}
   \label{fig:ACNabl}
\end{figure}

\subsection{Ablation Study}\label{AblationStudy}
In this section, we present the evaluation of the contribution of each network of the proposed framework. We also performed ablation studies to analyze the importance of each loss term. In addition, we tested the effects of pretraining and the iterative update algorithm proposed in Section~\ref{TrainingStrategy}. 
All experiments for the ablation studies were tested on the LIVE1 dataset and evaluated on the rate-distortion planes.

\subsubsection{Compact Representation Network}
To confirm the effect of the proposed CRNet, we experimented on the CRNet compared to the frameworks that simply downsample and restore \cite{bruckstein2003down, lin2006adaptive, wu2009low}. As illustrated in Fig.~\ref{fig:abl_CRnetcomp} (a)-(d), the CRNet outperformed the bicubic downsampling preprocessing at two scale factors: $0.5$ and $0.75$. When the scale factor was $0.5$, a higher performance improvement was obtained because the CRNet and PPNet have a larger capacity to compress and restore spatially as the scale factor decreases. 

Fig.~\ref{fig:abl_CRnetcomp} (e) and (f) exhibit the performance analysis according to the scale factor. A smaller scale factor result in a greater information loss for the original image; thus, the compression efficiency is improved only at a low bit rate. 
In contrast, in a high bit-rate environment, it is better to maintain the original scale.
An efficient scale factor exists according to the bit rate, suggesting that the scale factor should be determined adaptively according to the target rate.

\subsubsection{Auxiliary Codec Network}
To prove the superiority of the architecture of the proposed ACN, we conducted a performance comparison by imitating real codecs with several ACN structures. First, we compared the JPEG-based structure with the residual block-based CNN structure of the VCNN~\cite{zhao2019learning}. The results in Fig.~\ref{fig:abl_ACNmodelcomp} (a) and Table~\ref{tab:ACNmodel} reveal that JPEG-based ACN mimics JPEG decoded images better than VCNN. In particular, when Depth $N$ of the JPEG-based ACN is $12$, the imitation performance is over $40 dB$ in all QFs. The residual block-based CNN structure cannot follow the behavior of the JPEG codec regardless of the depth of the network.

In contrast, the proposed JPEG-based ACN generates a decoded image similar to the output of JPEG. The learned ACN expresses contouring and ringing artifacts, which are typical compression artifacts of JPEG. As the proposed ACN method closely follows the codec operation, backpropagation can be performed for the CRNet with a small error.

In addition, we conducted a comparative experiment on the HEVC-intra-based ACN. We compared the HEVC-based ACN with the VCNN and compared using the original image alone and with a prediction image for the input image of the HEVC-intra-based ACN. The results in Fig.~\ref{fig:abl_ACNmodelcomp} (b) and Table~\ref{tab:HEVCACN} indicate that the proposed HEVC-intra-based ACN structure has superior HEVC imitation performance compared with the VCNN with a Resblock-based CNN structure. Furthermore, we improved the ACN to take the prediction image as input with the original image, dramatically improving the imitation performance.

\begin{table*}
  \centering
  \caption{Quantitative Percent (\%) Error and Number of Parameters Comparison of Bits per Pixel (BPP) Estimation by Different Bit Estimation Network (BENet) Structures on Set14 and LIVE1 Datasets}
    \begin{tabular}{ccccccccccc}
    \toprule
    \toprule
    \multirow{2}[4]{*}{Structure} & \multicolumn{2}{c}{QF=10} & \multicolumn{2}{c}{QF=20} & \multicolumn{2}{c}{QF=40} & \multicolumn{2}{c}{QF=80} & \multicolumn{1}{c}{\multirow{2}[4]{*}{Average}} & \multicolumn{1}{c}{\multirow{2}[4]{*}{Number of Parameters}} \\
\cmidrule{2-9}      & Set14 & LIVE1 & Set14 & LIVE1 & Set14 & LIVE1 & Set14 & LIVE1 &   &  \\
    \midrule
    ResNet-18~\cite{he2016deep} & 1.598\% & \textbf{0.999}\% & 1.687\% & 1.535\% & 1.571\% & 1.591\% & \textbf{1.134}\% & \textbf{1.327}\% & 1.430\% & 15.199M \\
    ResNet-50~\cite{he2016deep} & 1.873\% & 1.934\% & 1.520\% & 1.364\% & \textbf{0.945}\% & \textbf{1.180}\% & 1.705\% & 2.005\% & 1.566\% & 29.287M \\
    BENet & \textbf{1.001}\% & 1.456\% & \textbf{1.221}\% & \textbf{1.305}\% & 1.072\% & 1.909\% & 1.211\% & 1.661\% & \textbf{1.355}\% & \textbf{2.098M} \\
    \bottomrule
    \bottomrule
    \end{tabular}%
  \label{tab:BENet}%
\end{table*}%

\subsubsection{Bit Estimation Network}
To prove the superiority of the structure of the proposed BENet, we compared it with ResNet~\cite{he2016deep}, a representative network architecture for regression. As a result of the experiment in Table~\ref{tab:BENet}, BENet demonstrated better bit prediction performance than ResNet. In particular, BENet is a more advantageous structure in that the number of parameters is relatively small.

\subsubsection{Simultaneous Learning Strategy}\label{exp:Learningmethod}
We conducted a comparative experiment with the learning strategies of recent CRNet-based papers~\cite{jiang2017end, zhao2019learning}. In practice, the performance of the gradient backpropagation should be compared to know how well the learning strategies mimic the codec role. However, real nondifferentiable codecs have no ground truth for gradient propagation. Therefore, we analyzed the mimicking ability of the proposed method through the final compression performance after optimizing CRNet and PPNet.

In Fig.~\ref{fig:ACNabl}, we compared the compression methods using algorithm preprocessing and postprocessing networks. The methodologies selected for comparison are as follows: first, the method for learning the CRNet and PPNet alternately by approximating the codec as an identity function as in~\cite{jiang2017end} (green line) and using the VCNN~\cite{zhao2019learning} (magenta line), and second, the method for simultaneous optimization by directly connecting the CRNet and PPNet (cyan line), using the VCNN (purple line) and proposed JPEG-based ACN (blue line). Furthermore, we additionally performed alternate learning of the CRNet and PPNet with the JPEG-based ACN (red line) to compare the influence of alternate learning and simultaneous learning. For a fair comparison, all experiments used the fixed network structure of CRNet and PPNet, and the same training database as described in Section~\ref{Setting}.

In Fig.~\ref{fig:ACNabl} (a), the proposed JPEG-based ACN with simultaneous learning outperforms the other methods. In learning the CRNet and PPNet alternately by approximating the codec as an identity function, the codec characteristics are repeatedly reflected in the PPNet to demonstrate good performance. However, an error occurs because the codec is assumed to be an identity function when learning the CRNet. Additionally, limitations exist in the VCNN, as it is difficult to sufficiently transfer the codec characteristics to the CRNet because of the poor approximation of the codec.
 
In the case of simultaneous learning by directly connecting the CRNet and PPNet and with the Resblock-based VCNN, the compression performances are significantly worse than the others. Table~\ref{tab:ACNmodel} demonstrates that the VCNN is closer to the JPEG decoded image than the original image (identity function). However, as illustrated in the visual comparison in Fig.~\ref{fig:abl_ACNmodelcomp} (a), compression mimicking images generated from the VCNN have almost no observed JPEG compression noise patterns and unpredictable noise. End-to-end learning with the VCNN, which is updated through iterative learning and is changeable, unlike an identity function that generates a constant value, leads to worse optimization. Therefore, the VCNN can generate even greater error propagation than the identity function approximation.
 
In alternate learning, PPNet is continuously trained from a decoded image obtained from the real codec to compensate for approximation errors. However, for simultaneous learning, if a sufficient codec approximation is not satisfied, error propagation continues to CRNet and PPNet during training, resulting in performance degradation. The JPEG-based ACN exhibited better performance in simultaneous learning than in alternating learning. This result reveals that the proposed learning method, in which CRNet and PPNet are optimized simultaneously, is more effective if good mimicking performance is guaranteed. In Fig.~\ref{fig:ACNabl} (b), the experimental results demonstrated that, as the imitation performance gradually decreased with a small value of $N$, the overall compression performance also gradually decreased. The experiments indicate that the imitation performance is proportional to the overall compression performance and that better mimicking performance improves the compression performance through a more precise optimization of the preprocessing network.

\begin{figure}
   \centering
   \subfloat[][]{\includegraphics[width=.5\linewidth]{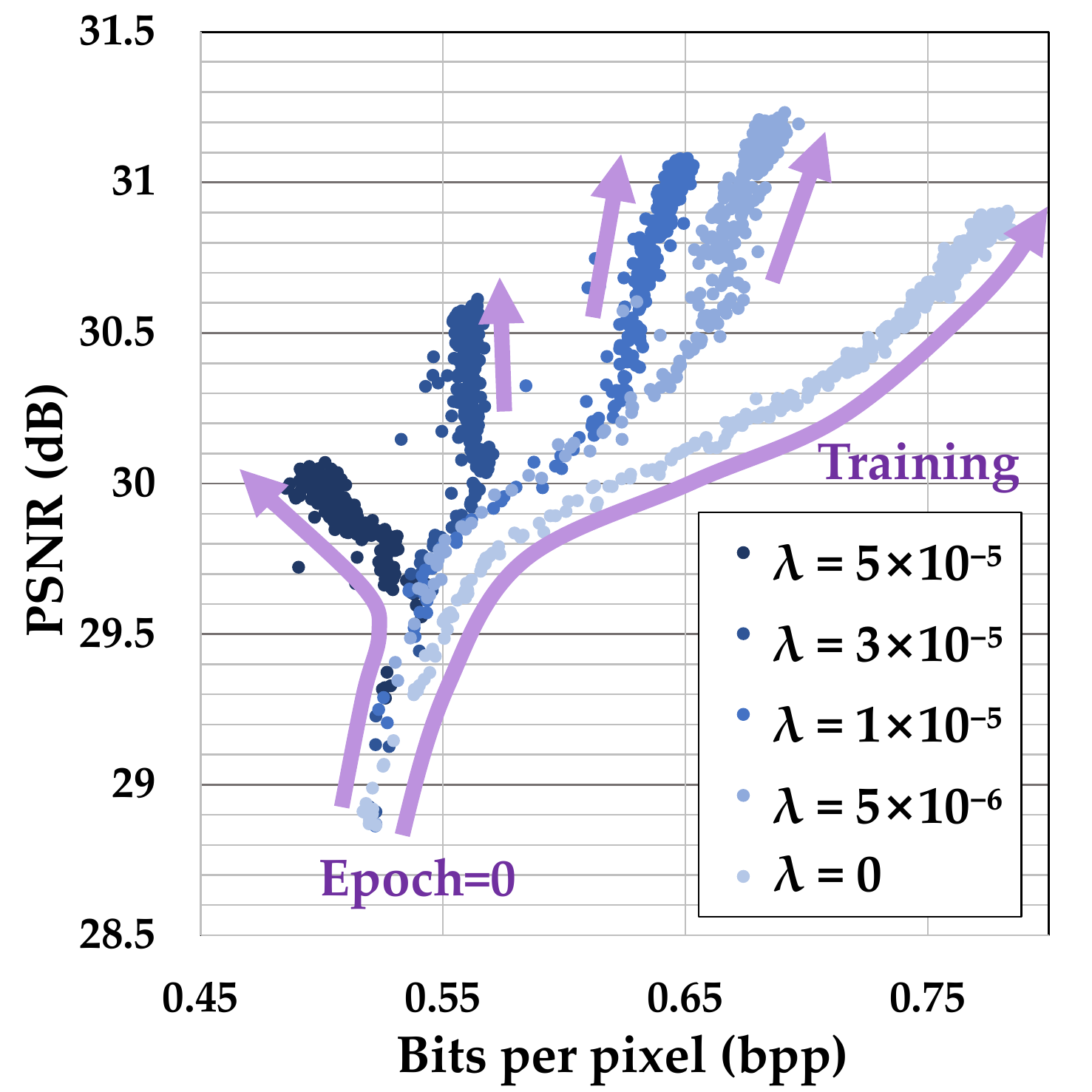}}
   \subfloat[][]{\includegraphics[width=.5\linewidth]{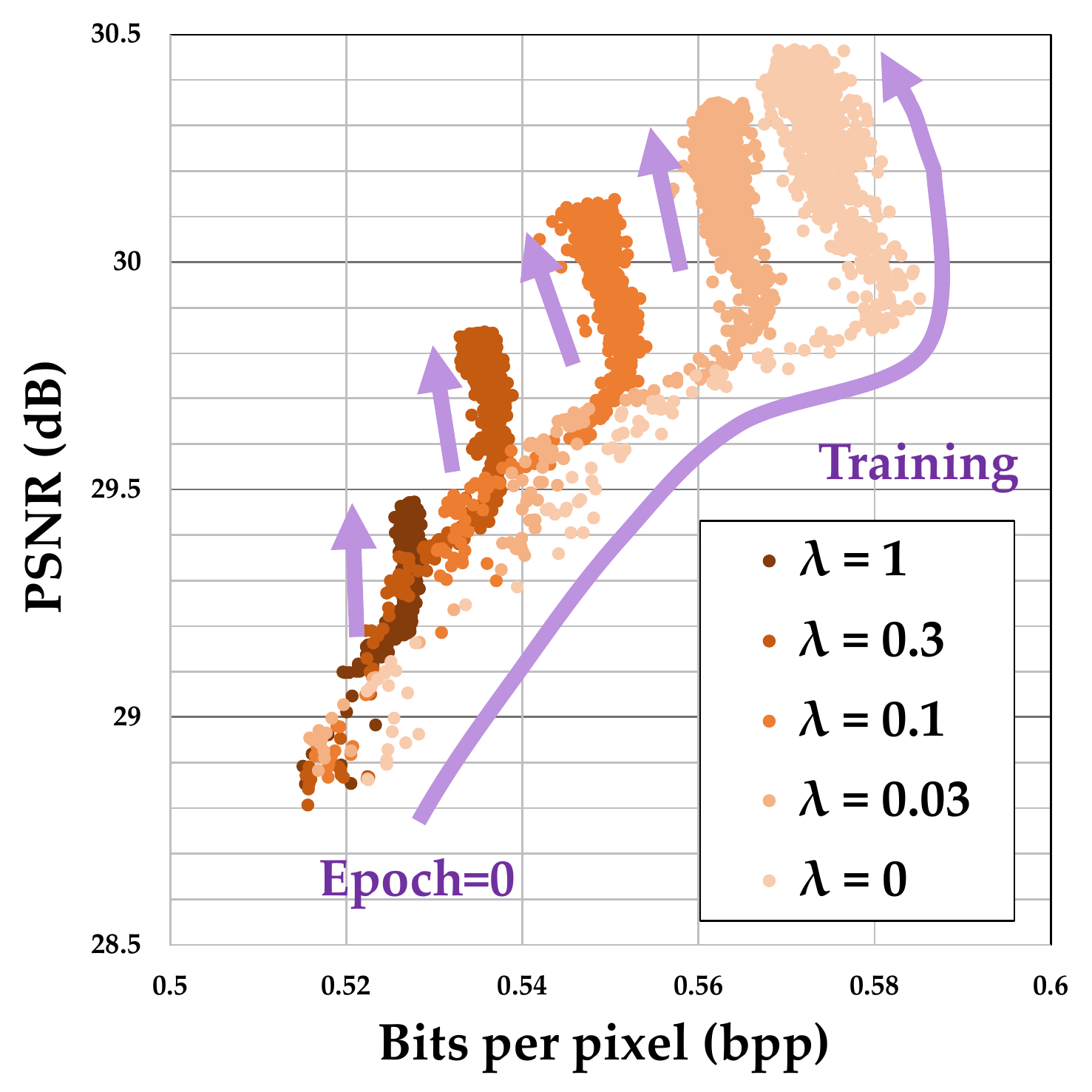}}\
   \subfloat[][]{\includegraphics[width=.5\linewidth]{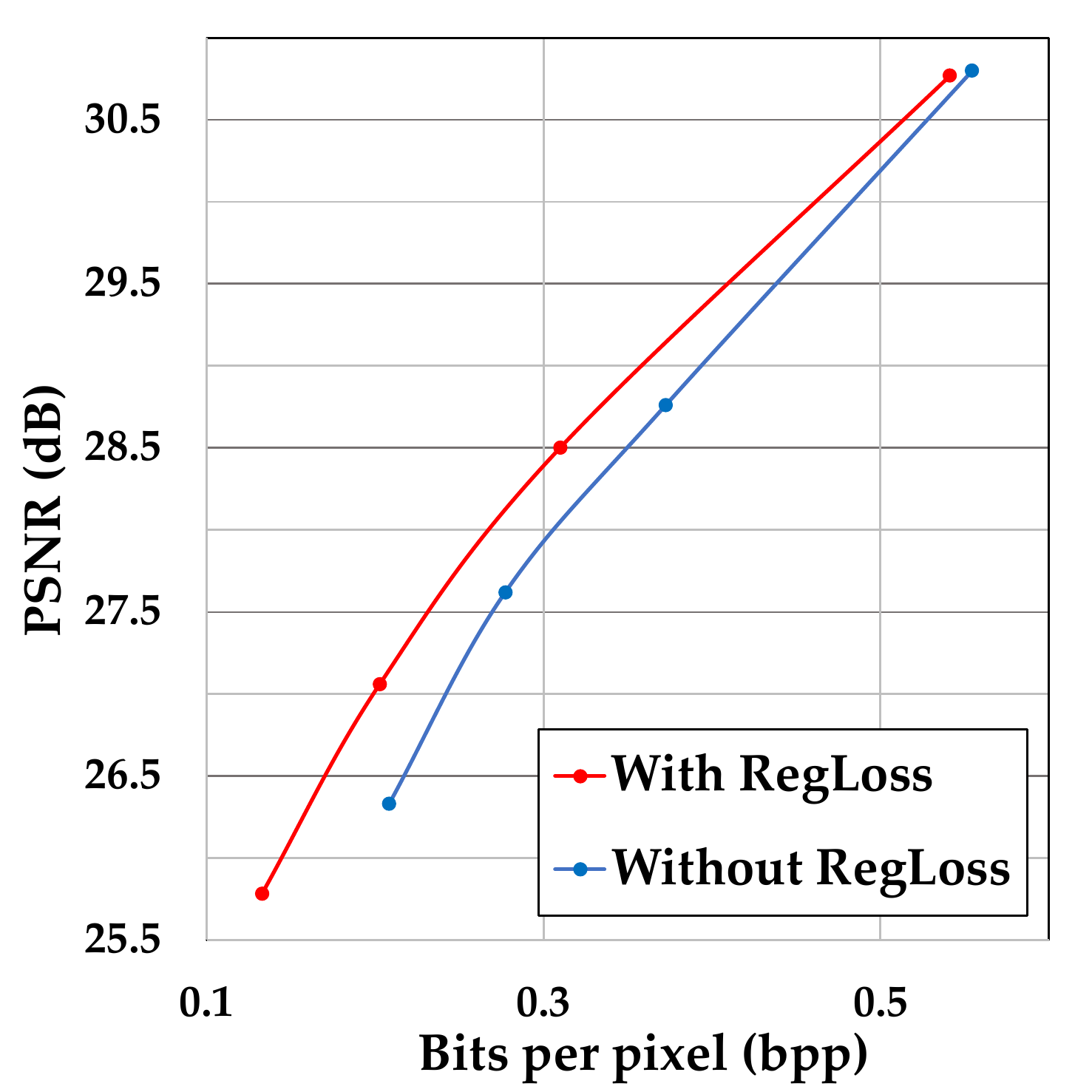}}
   \caption{Rate-distortion performance comparison according to the weight of the loss function. Rate-distortion point of compressed image (JPEG $QF=80$) with increasing learning epochs according to (a) the bit loss $\lambda_{bit}$, and (b) regularization loss $\lambda_{reg}$, and (c) rate-distortion performance comparison with $\lambda_{reg} = 0.1$ or without regularization loss $\lambda_{reg}$ after the training process is complete.}
   \label{fig:abl_loss}
\end{figure}

\begin{figure}
   \centering
   \subfloat[][]{\includegraphics[width=.5\linewidth]{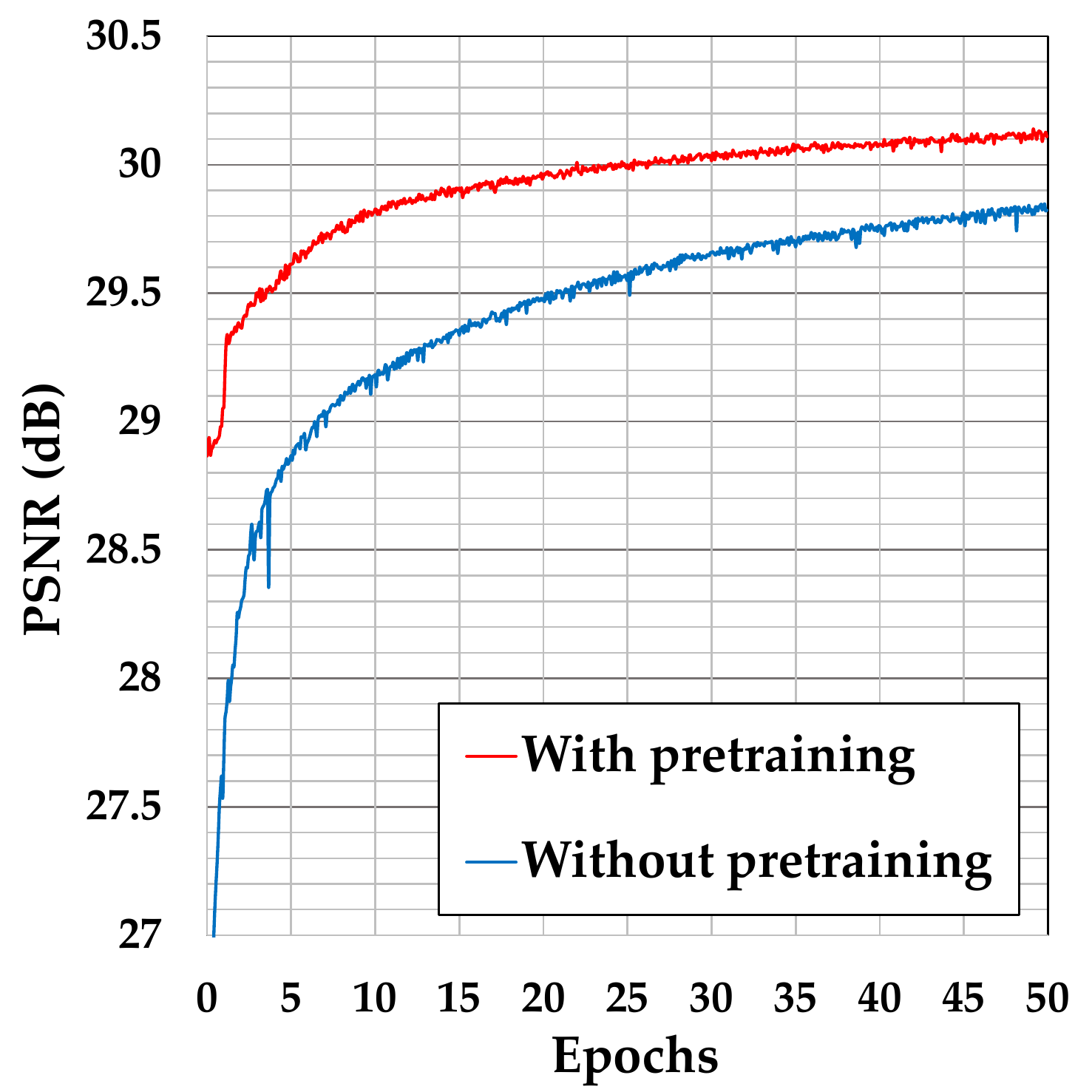}}
   \subfloat[][]{\includegraphics[width=.5\linewidth]{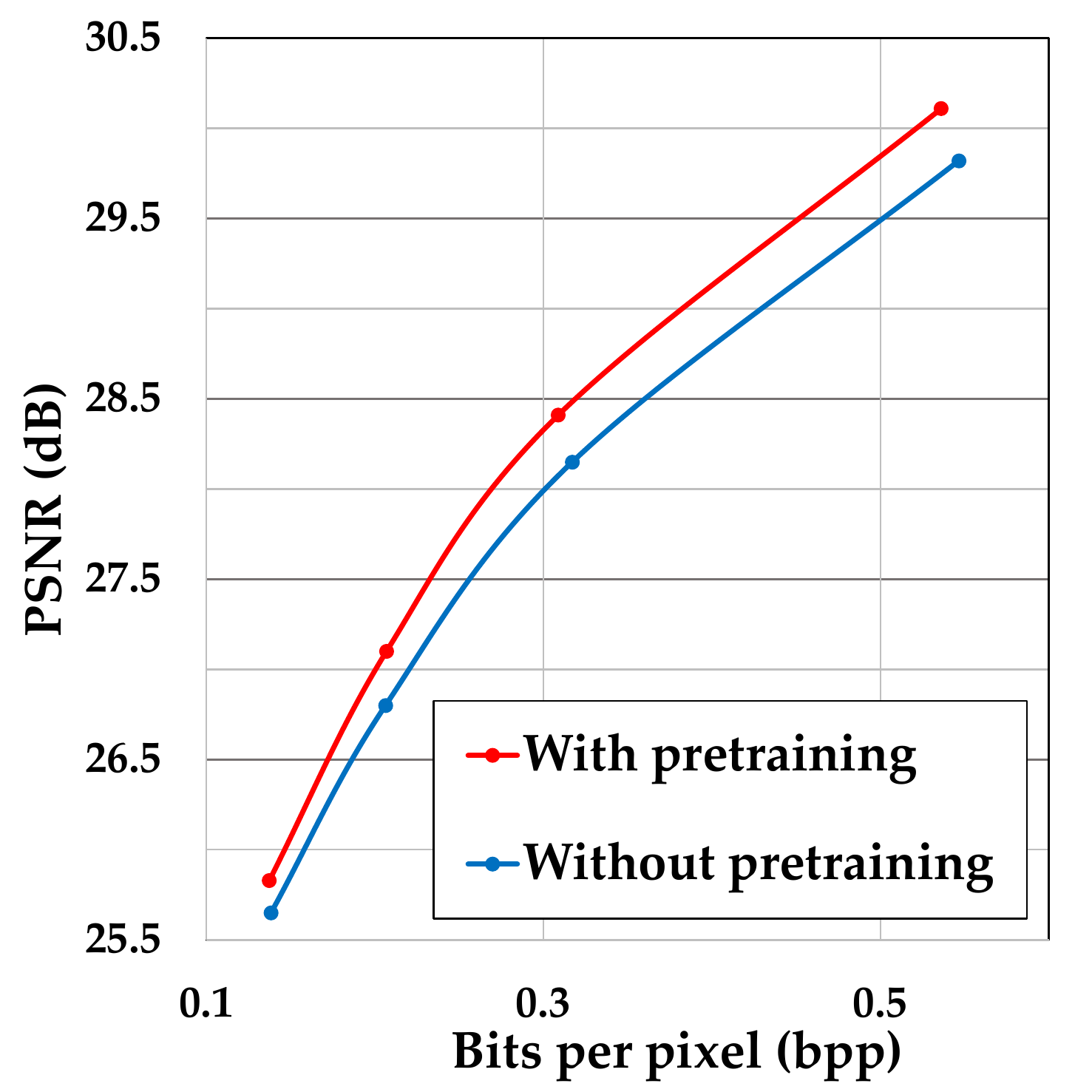}}
   \caption{Performance comparison with and without pretraining. (a) Change in the peak signal-to-noise ratio (PSNR) between the original and reconstructed images (JPEG $QF=80$) in the end-to-end model according to the training epochs and (b) comparison of the rate-distortion performance after the training process is complete.}
   \label{fig:abl_pretraining}
\end{figure}

\begin{figure}
   \centering
   \subfloat[][]{\includegraphics[width=.5\linewidth]{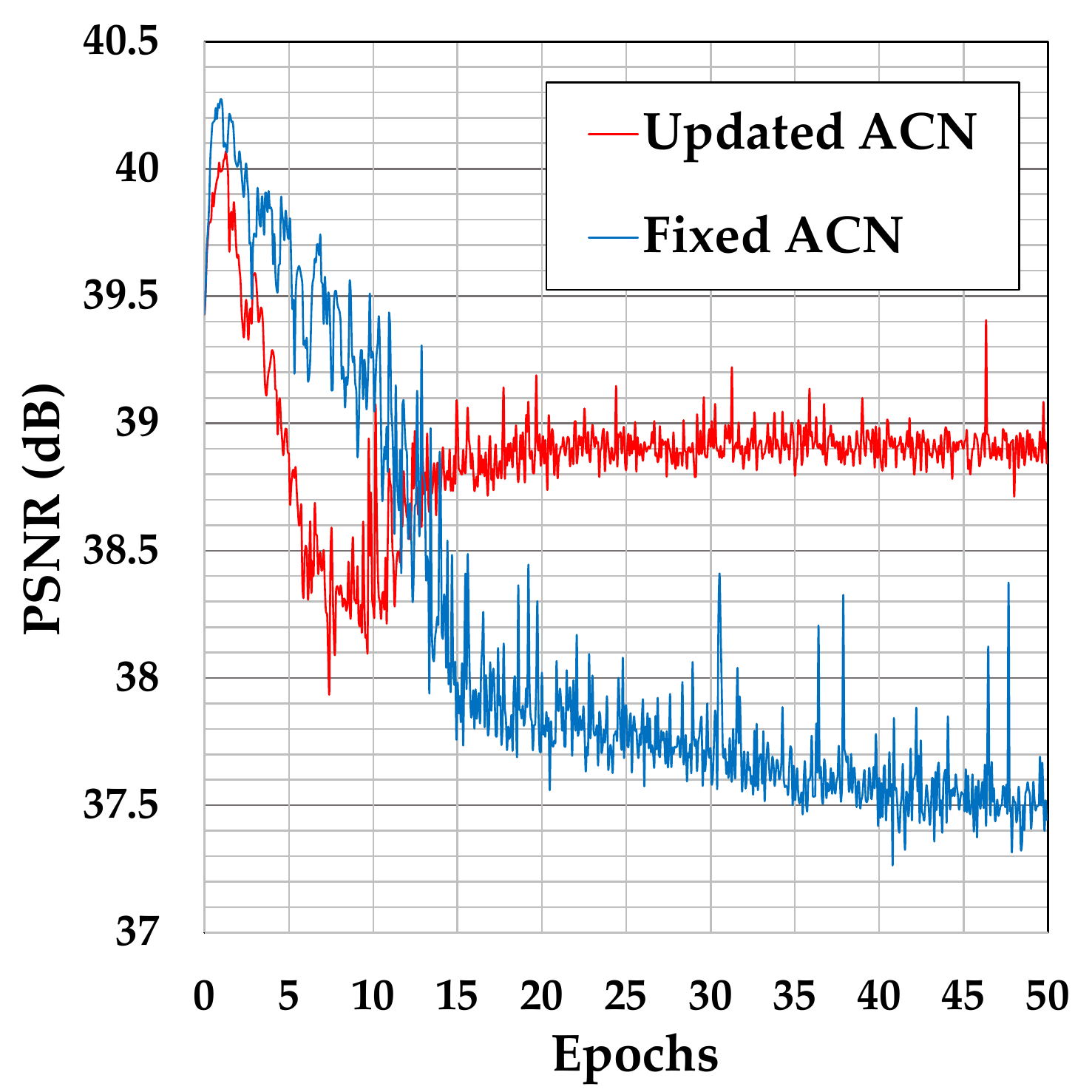}}
   \subfloat[][]{\includegraphics[width=.5\linewidth]{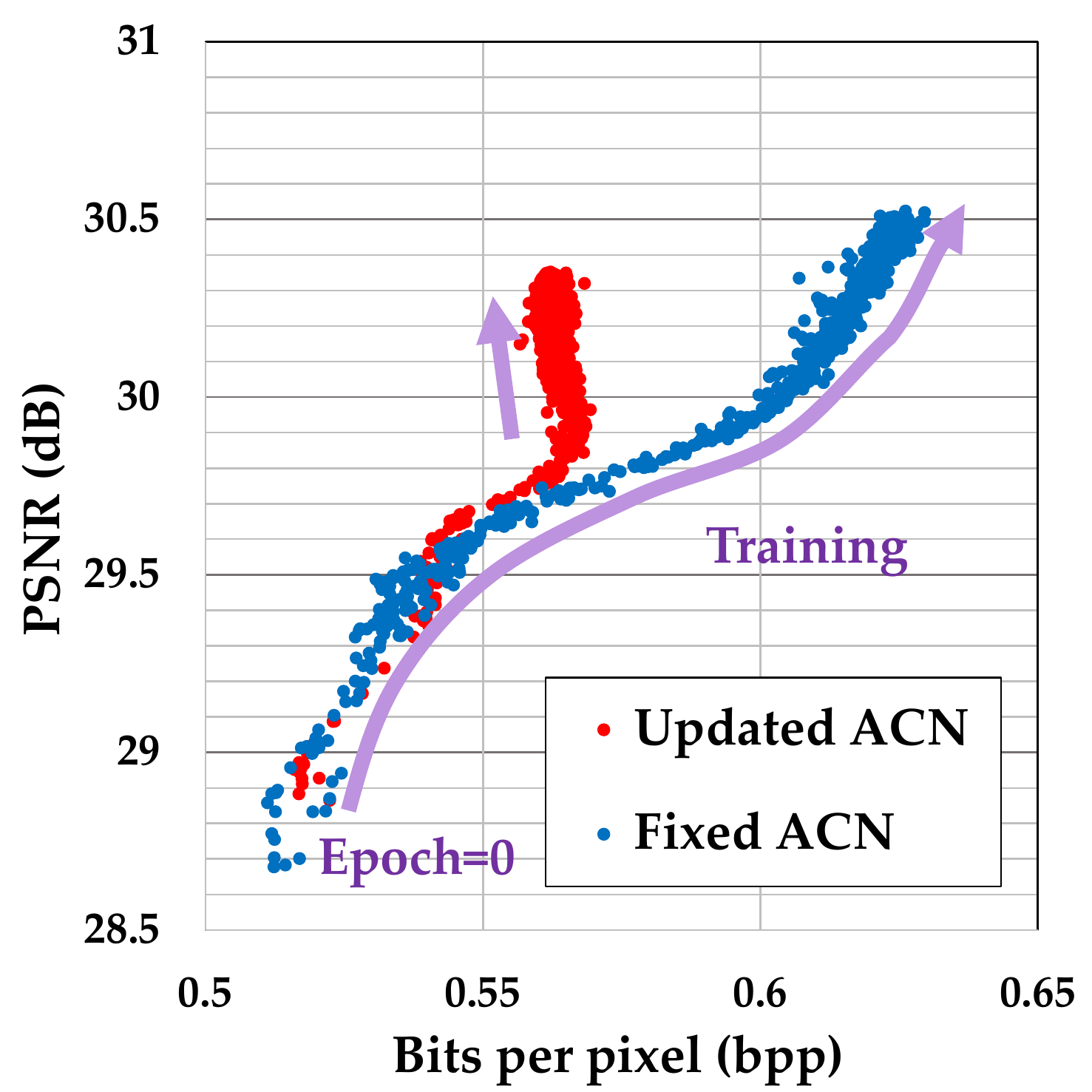}}
   \caption{Performance comparison with and without an iterative update in the simultaneous learning process. (a) Change in the peak signal-to-noise ratio (PSNR) between the output image generated from the JPEG-based auxiliary codec network (ACN) and the decoded image from the JPEG decoder ($QF=80$) according to training epochs and (b) change in the rate-distortion point as training progresses.}
   \label{fig:abl_Iterative}
\end{figure}

\subsubsection{Bit and Regularization Loss}\label{Exp:Loss}
The rate-distortion performance was evaluated according to the two loss weights, $\lambda_{bit}$ and $\lambda_{reg}$, to determine the effectiveness of the bit and regularization loss. According to the training progress, the rate-distortion performance results are expressed as traces according to the training epoch to analyze the change in performance. As displayed in Fig.~\ref{fig:abl_loss} (a), when learning the network without considering the bit loss ($\lambda_{bit}=0$), the final reconstructed image becomes closer to the original image, but the number of bits generated during compression increases significantly, resulting in poor coding efficiency. In contrast, using a proper $\lambda_{bit}$ prevents these problems.
Fig.~\ref{fig:abl_loss} (b) indicates that the network has a better training procedure with stable convergence with the regularization loss by preserving the structure of natural images. 
If the influence of the regularization loss increases, the learning of the CRNet is restricted. 
In this case, no significant change in performance exists from the initial state. 
Fig.~\ref{fig:abl_loss} (c) indicates that the coding efficiency is better when learning with the regularization loss ($\lambda_{reg}=0.1$) than learning without the regularization loss. A standard codec is designed to compress natural images; thus, regularization loss prevents a decrease in coding efficiency from compressing images with unnatural patterns.

\subsubsection{Pretraining Strategy}\label{Exp:Pretraining}
We analyzed the effect of the pretraining strategy as described in Section~\ref{pretraining}. Fig.~\ref{fig:abl_pretraining} presents the performance comparison according to whether pretraining occurred. The experimental results reveal faster convergence with better coding efficiency when performing pretraining on the CRNet and PPNet. The absence of pretraining means that the parameters of the CRNet and PPNet are initialized with random values. In this case, it is challenging to converge in the desired direction because the ACN and BENet do not work properly at the beginning of the training process.

\subsubsection{Iterative Updating Strategy}
The iterative updating process is proposed in Section~\ref{iteratvielearning} to reduce errors due to the approximation functions. The fixed ACN is compared with the repeatedly updated ACN to demonstrate the effectiveness of the proposed training scheme. 
Fig.~\ref{fig:abl_Iterative} (a) depicts the comparison of the codec imitation performance according to the training epochs, and Fig.~\ref{fig:abl_Iterative} (b) displays the change in coding efficiency as training progresses. When the ACN is fixed, the imitation performance of the ACN decreases gradually as the epoch increases. In contrast, when the ACN is continuously updated, the imitation performance does not deteriorate, and it converges. Guaranteeing the performance of the ACN helps learning to increase the coding efficiency of the entire framework with less approximation error.

\subsection{Comparison with State-of-the-art Methods}

\begin{figure}
    \centering
    \subfloat[][]{\includegraphics[width=.5\linewidth]{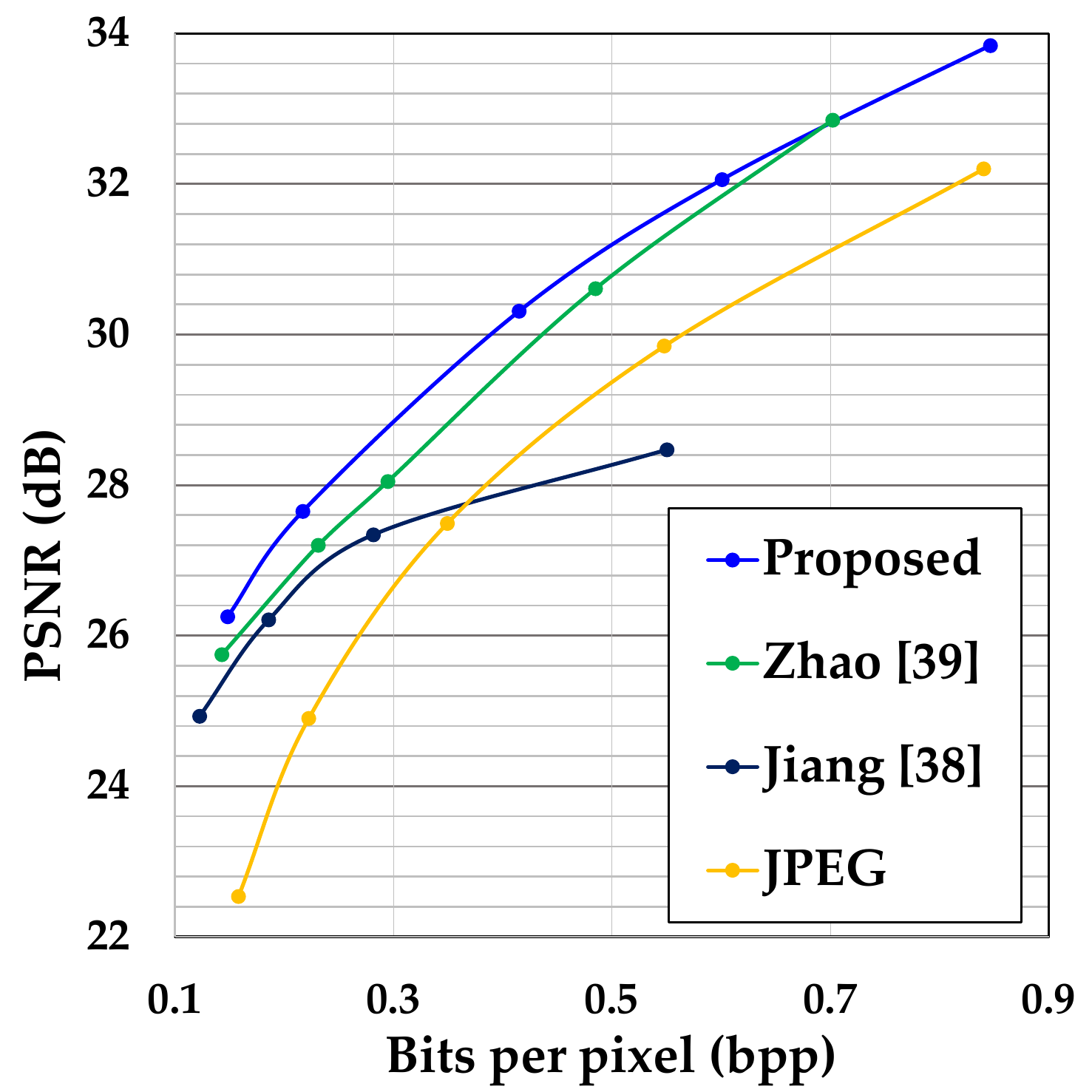}}
    \subfloat[][]{\includegraphics[width=.5\linewidth]{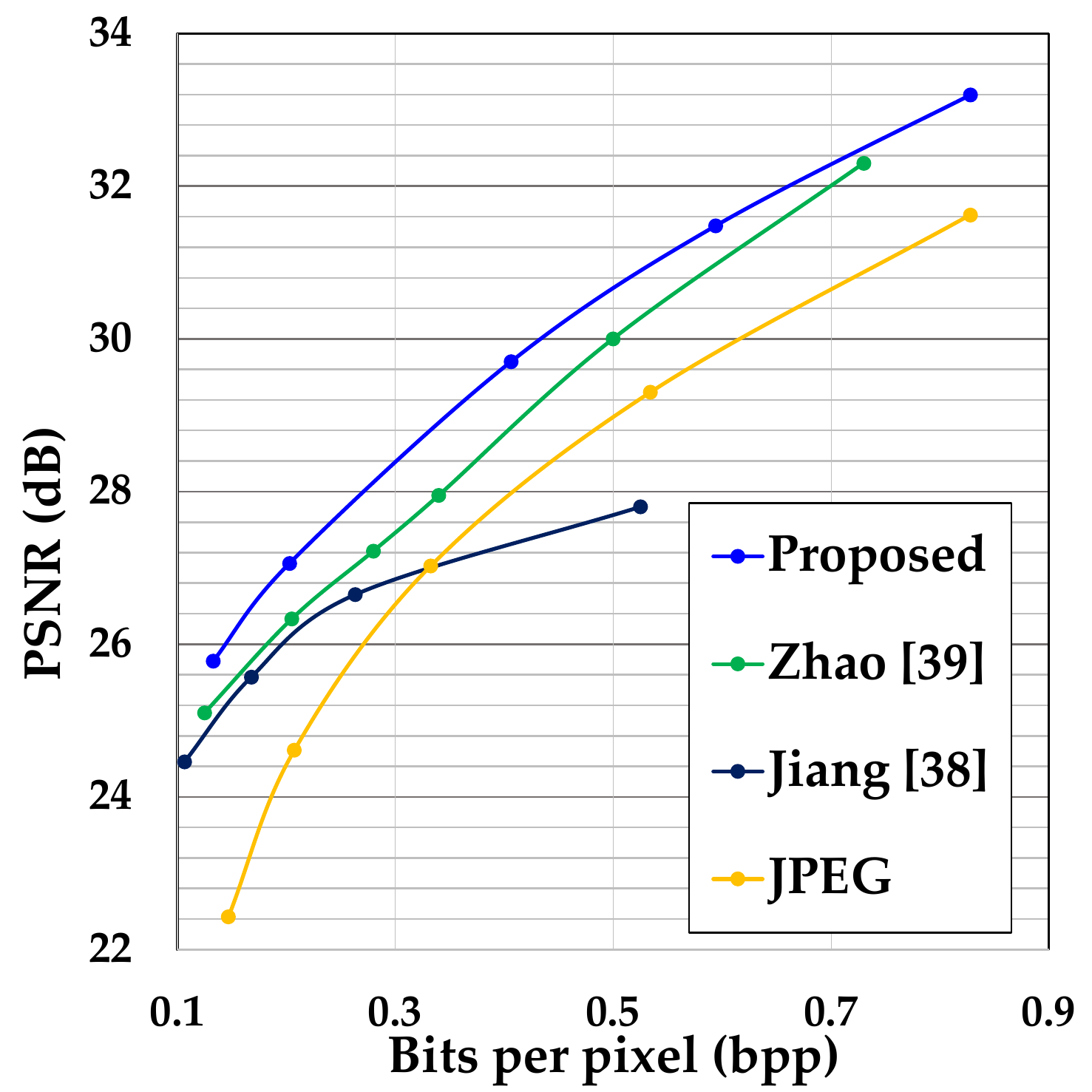}}\\
    \subfloat[][]{\includegraphics[width=.5\linewidth]{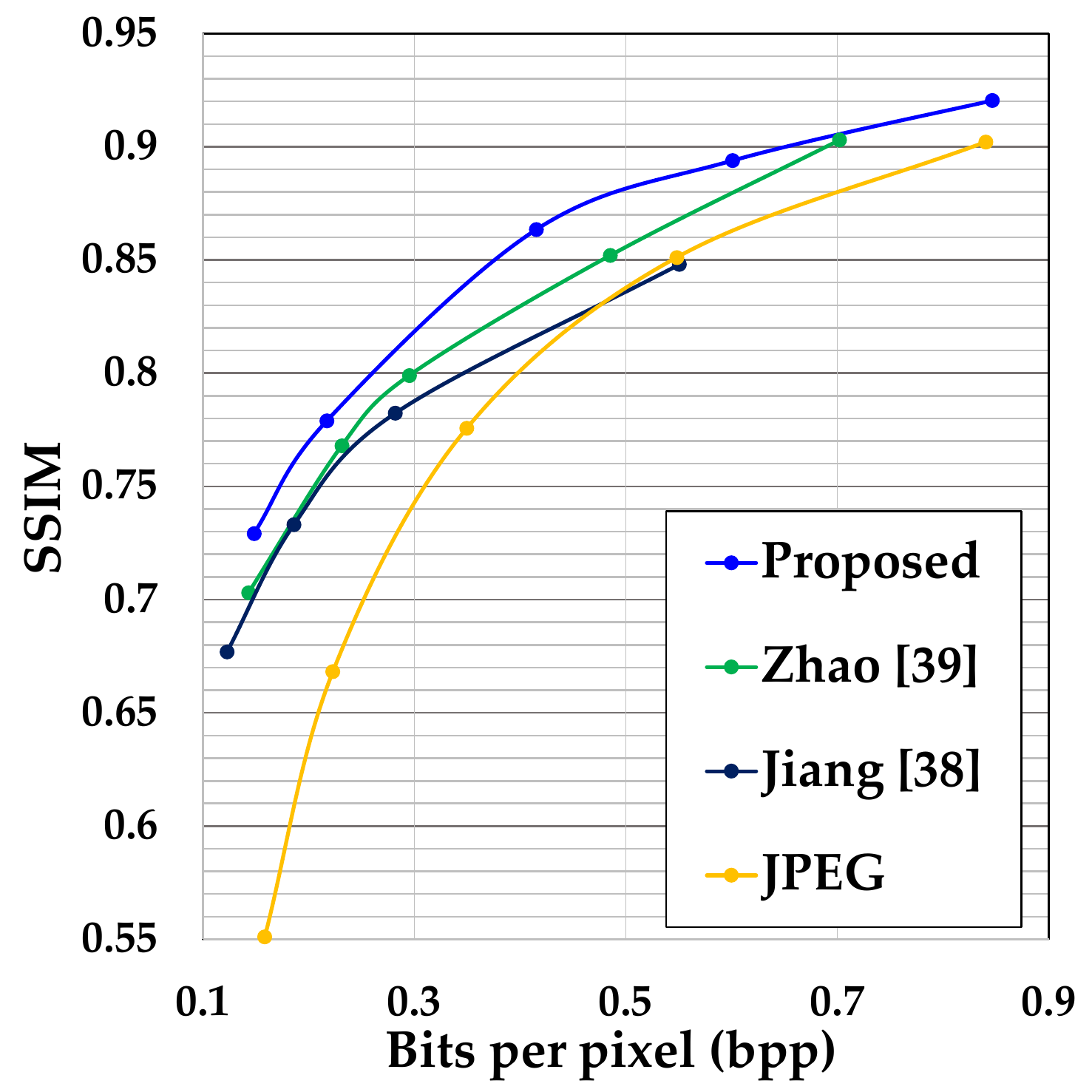}}
    \subfloat[][]{\includegraphics[width=.5\linewidth]{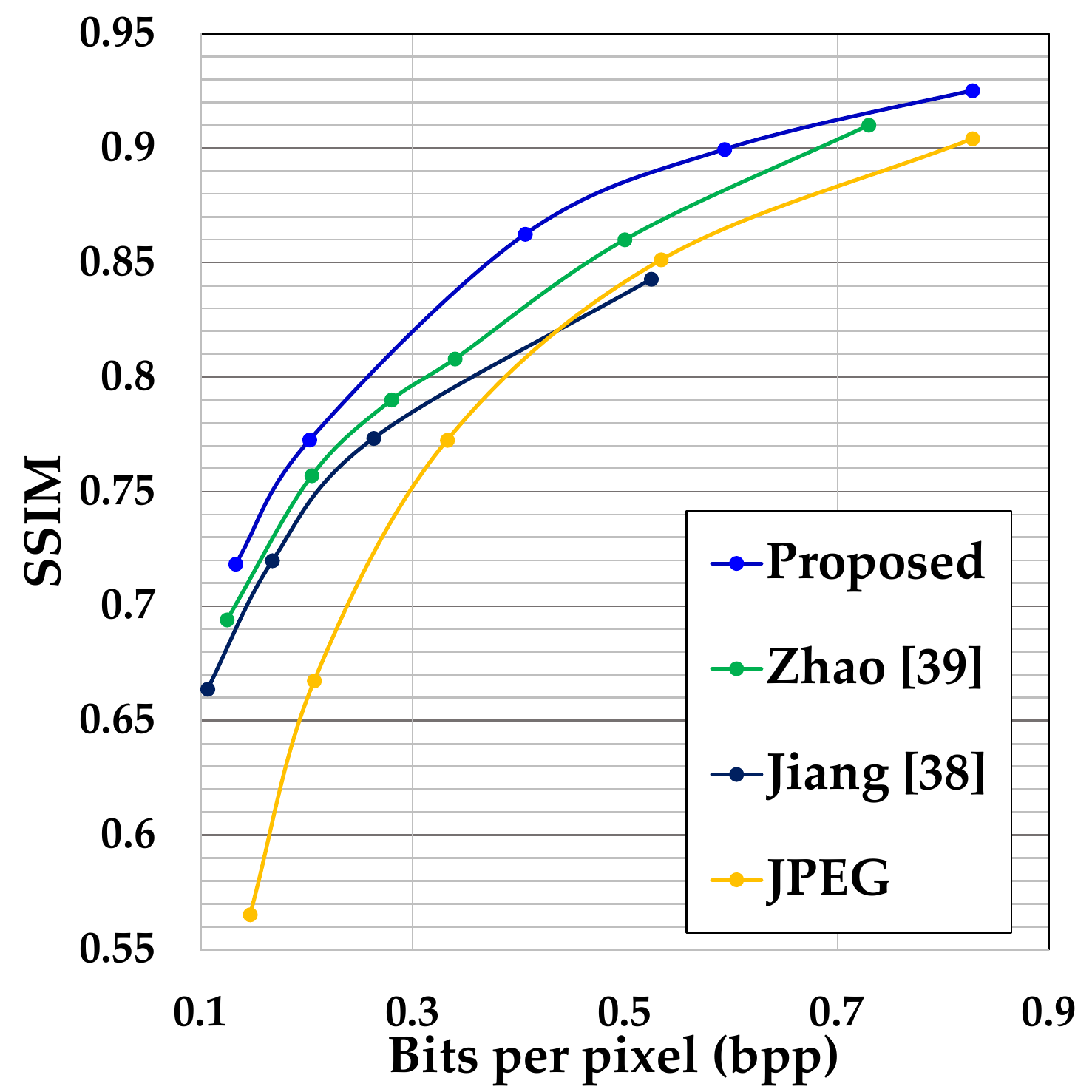}}
	\caption{Rate-distortion performance per peak signal-to-noise ratio (PSNR) and structural similarity index measure (SSIM) of different compression algorithms on test image datasets: (a), (c) Set14 and (b), (d) LIVE1.}
	\label{fig:exp_JPEG}
\end{figure} 

\begin{figure}
   \centering
   \subfloat[][]{\includegraphics[width=.5\linewidth]{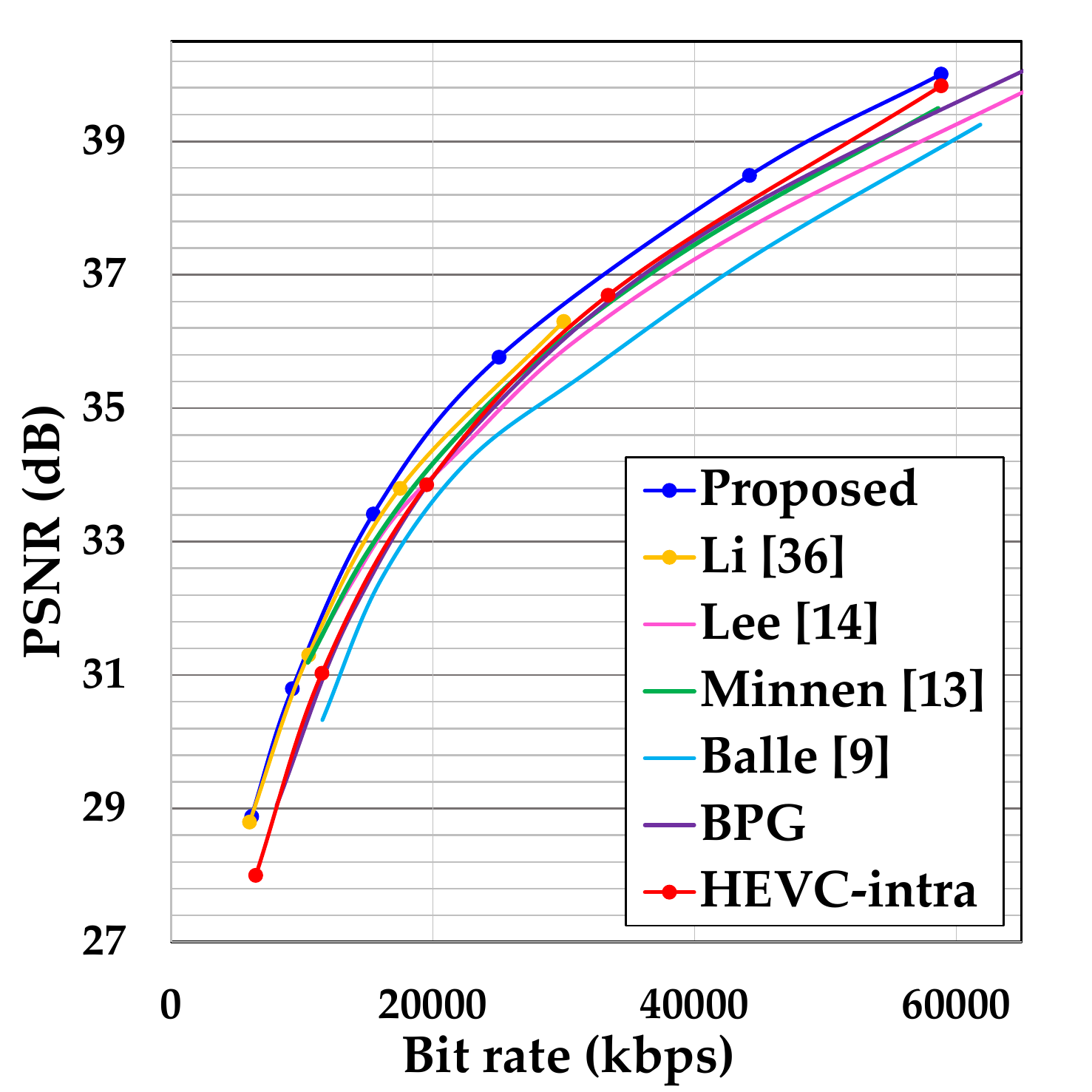}}
   \subfloat[][]{\includegraphics[width=.5\linewidth]{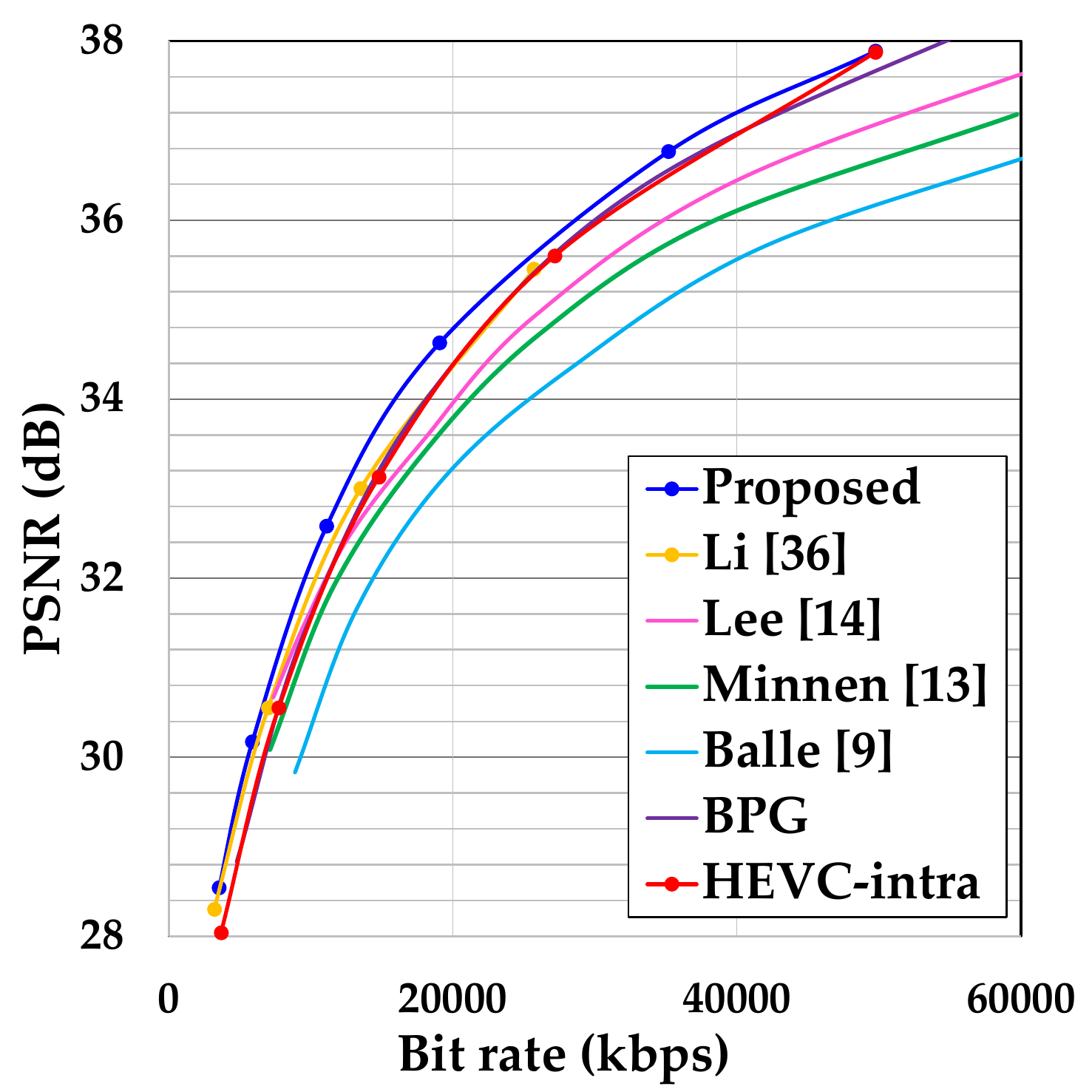}}\\
   \subfloat[][]{\includegraphics[width=.5\linewidth]{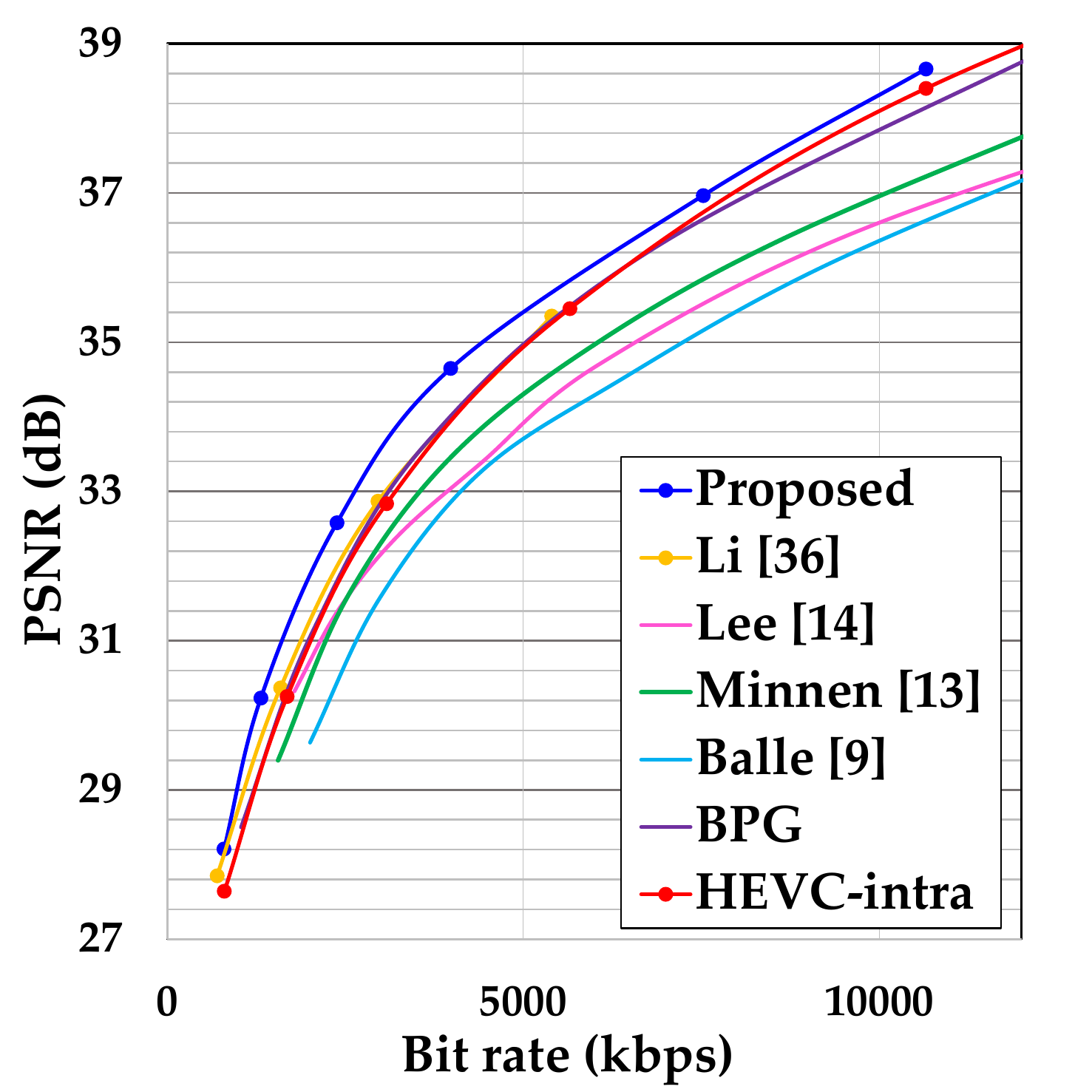}}
   \subfloat[][]{\includegraphics[width=.5\linewidth]{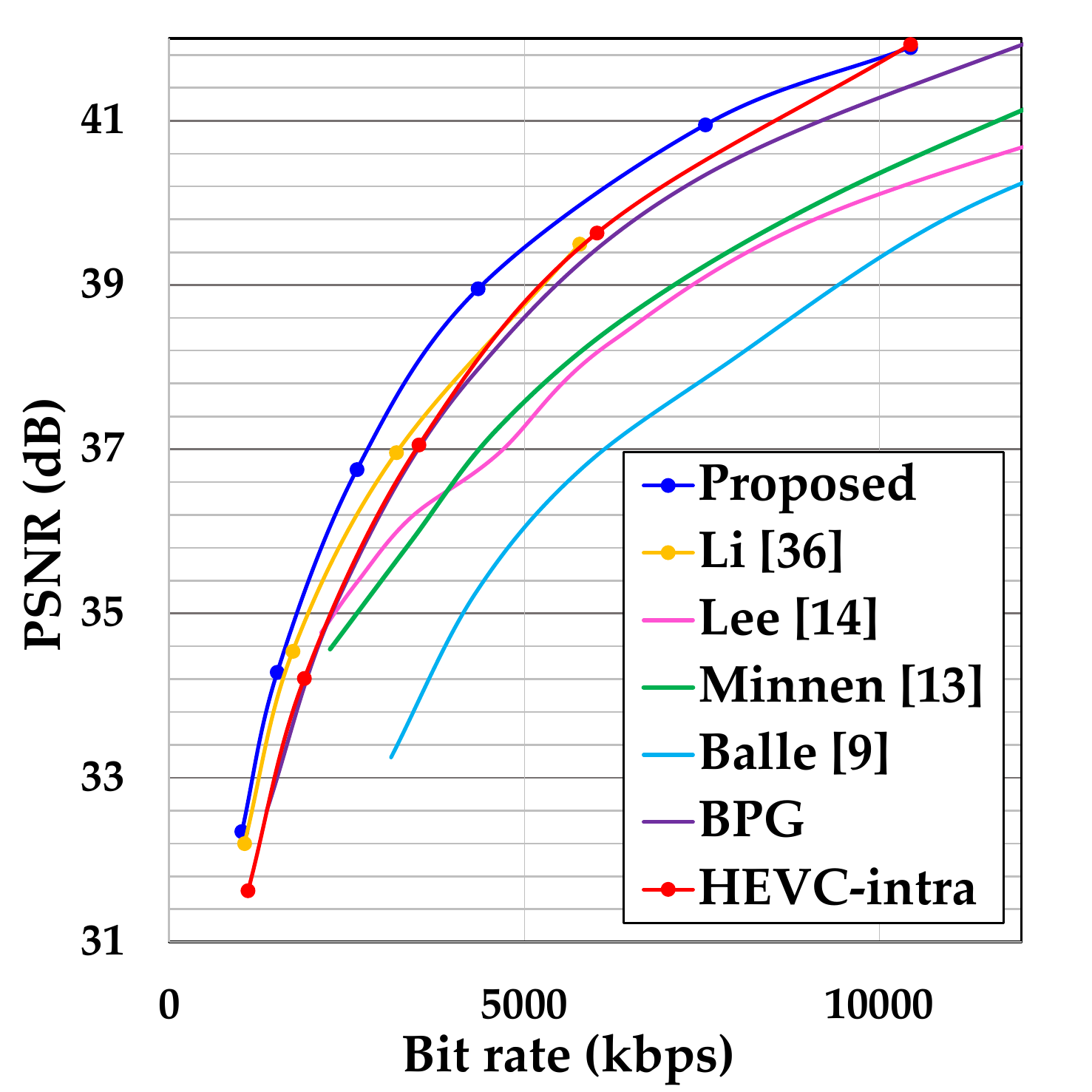}}
   \caption{Rate-distortion curves of several HEVC test sequences~\cite{bossen2013common}: (a) PeopleOnStreet $(2160\times1440)$, (b) Cactus $(1920\times1080)$, (c) BasketballDrill $(832\times480)$, and (d) Johnny $(1280\times720)$.}
   \label{fig:exp_HEVC}
\end{figure}

\subsubsection{JPEG-based Model}
We compared the coding efficiency of the proposed JPEG-based ACN with depth $N$ of $12$ with JPEG and other standard compatible compact representation frameworks, such as those by Jiang~\cite{jiang2017end} and Zhao~\cite{zhao2019learning}. We used the experimental results described in \cite{zhao2019learning}; however, in \cite{jiang2017end}, the compression rate and distortion results for high QF are not described, so we newly trained the model from \cite{jiang2017end} and used these results.

Considering that the CRNet is a useful tool under a low bit-rate environment, the QF of JPEG was set from $2$ to $40$ for comparison. In the proposed method, we adaptively determined the scale factor of the CRNet and PPNet according to the bit rate. The scale factors are determined to be 0.5, 0.75, and 1, respectively in three ranges of bit rate from low to high.

As depicted in Fig.~\ref{fig:exp_JPEG}, the proposed method outperforms the methods by Jiang and Zhao in all bits per pixel (bpp) ranges. Moreover, the proposed method exhibits the best performance on all test datasets. The difference between the proposed algorithm and the existing method is the learning method for the CRNet. From the experimental results, the well-trained CRNet improves compression performance. 

\begin{table*}
  \centering
  \caption{BD-Rate Comparison of Standard Compatible Frameworks on HEVC common test conditions~\cite{bossen2013common} with All Intra Main configurations}
    \begin{tabular}{cccccccc}
    \toprule
    \toprule
      &   & \multicolumn{2}{c}{Li (Frame level) \cite{li2018learning}} & \multicolumn{2}{c}{Li (Block level) \cite{li2018learning}} & \multicolumn{2}{c}{Proposed} \\
    \midrule
      &   & \hspace{0.3cm}PSNR\hspace{0.3cm} & \hspace{0.3cm}SSIM\hspace{0.3cm} & \hspace{0.3cm}PSNR\hspace{0.3cm} & \hspace{0.3cm}SSIM\hspace{0.3cm} & \hspace{0.3cm}PSNR\hspace{0.3cm} & \hspace{0.3cm}SSIM\hspace{0.3cm} \\
    \midrule
    \multicolumn{1}{c}{\multirow{2}[0]{*}{Class A}} & Traffic & -12.4\% & -17.1\% & -11.8\% & -14.8\% & -15.0\% & -22.4\% \\
      & PeopleOnStreet & -9.5\% & -14.4\% & -11.7\% & -14.5\% & -15.4\% & -23.4\% \\
    \midrule
    \multicolumn{1}{c}{\multirow{5}[0]{*}{Class B}} & Kimono & -13.0\% & -14.5\% & -9.0\% & -10.8\% & -11.7\% & -20.6\% \\
      & ParkScene & -8.8\% & -15.5\% & -8.3\% & -13.0\% & -11.6\% & -19.4\% \\
      & Cactus & -7.1\% & -20.0\% & -8.5\% & -12.9\% & -14.9\% & -22.6\% \\
      & BasketballDrive & 7.0\% & -19.7\% & -8.0\% & -11.3\% & -11.3\% & -23.8\% \\
      & BQTerrace & 6.2\% & -20.8\% & -4.8\% & -13.0\% & -14.1\% & -20.7\% \\
    \midrule
    \multicolumn{1}{c}{\multirow{4}[0]{*}{Class C}} & BasketballDrill & -10.7\% & -24.8\% & -7.5\% & -10.5\% & -19.2\% & -25.0\% \\
      & BQMall & 17.2\% & -23.4\% & -3.9\% & -8.0\% & -7.3\% & -19.9\% \\
      & PartyScene & 8.9\% & -26.2\% & -1.9\% & -6.5\% & -9.3\% & -20.2\% \\
      & RaceHorcesC & -6.8\% & -18.5\% & -8.2\% & -13.0\% & -14.2\% & -19.1\% \\
    \midrule
    \multicolumn{1}{c}{\multirow{4}[0]{*}{Class D}} & BasketballPass & 6.5\% & -22.0\% & -4.6\% & -9.1\% & -20.1\% & -24.4\% \\
      & BQSquare & 7.8\% & -26.6\% & -1.8\% & -3.6\% & -15.4\% & -22.8\% \\
      & BlowingBubbles & 3.8\% & -18.8\% & -4.2\% & -8.9\% & -19.7\% & -24.1\% \\
      & RaceHorsesD & -13.5\% & -17.7\% & -13.0\% & -18.0\% & -18.5\% & -21.2\% \\
    \midrule
    \multicolumn{1}{c}{\multirow{3}[0]{*}{Class E}} & FourPeople & -3.9\% & -18.3\% & -9.1\% & -14.5\% & -15.2\% & -26.1\% \\
      & Johnny & -8.1\% & -12.9\% & -10.2\% & -11.8\% & -20.9\% & -28.9\% \\
      & KristenAndSara & -0.7\% & -20.0\% & -7.9\% & -14.0\% & -15.7\% & -22.8\% \\
    \midrule
    \multicolumn{1}{c}{\multirow{5}[0]{*}{Summary}} & Class A & -11.0\% & -15.8\% & -11.8\% & -8.7\% & -15.2\% & -22.9\% \\
      & Class B & -3.1\% & -18.1\% & -7.7\% & -13.1\% & -12.7\% & -21.4\% \\
      & Class C & 2.2\% & -23.2\% & -5.4\% & -11.5\% & -12.5\% & -21.1\% \\
      & Class D & 1.2\% & -21.3\% & -5.9\% & -14.8\% & -18.4\% & -23.1\% \\
      & Class E & -4.2\% & -17.1\% & -9.1\% & -11.9\% & -17.2\% & -25.9\% \\
    \midrule
    \multicolumn{2}{c}{Overall} & -3.0\% & -19.1\% & -8.0\% & -12.0\% & -15.2\% & -22.9\% \\
    \bottomrule
    \bottomrule
    \end{tabular}%
  \label{tab:exp_HEVC}%
\end{table*}%

\subsubsection{HEVC-intra-based Model}
The proposed method was compared with the conventional codecs, learnable codecs, and the competing standard compatible algorithm to apply the proposed method to the recent standard codec and exhibit state-of-the-art performance. For conventional codecs, the BPG image codec was also selected. The image compression frameworks proposed by Ball{\'e}~\cite{balle2018variational}, Minnen~\cite{minnen2018joint}, and Lee~\cite{lee2018context} were selected for the learnable codec, and the work by Li~\cite{li2018learning} was selected as the competing standard compatible algorithm. The rate-distortion performance comparison of the typical sequences of each class is illustrated in Fig.~\ref{fig:exp_HEVC}.

The performance of the learnable codec algorithms does not exceed that of the HEVC in the test sequences. In contrast, the standard compatible frameworks are proposed to boost the performance of the existing codec, exhibiting better coding efficiency than the other frameworks. In particular, our method reached state-of-the-art performance.

For a detailed performance comparison between the standard compatible frameworks, the BD-rate performance was compared in the HEVC test sequences. Considering that the standard compatible framework is effective in a low bit-rate environment, the QP of the reference HEVC was set to $32$, $37$, $42$, and $47$. The BD-rate results are summarized in Table~\ref{tab:exp_HEVC}. The results reveal that the proposed method achieves an average BD-rate reduction of $15.2\%$ in all classes on the PSNR metric and $22.9\%$ on the SSIM metric. The proposed method also outperforms Li's frame-level and block-level scheme \cite{li2018learning} for both the PSNR and SSIM metrics. In addition, \cite{li2018learning} assumed that the standard codec is an identity function, similar to that in \cite{jiang2017end}; therefore, the CRNet and PPNet are directly connected, and the CRNet is learned through joint learning of the entire network. The proposed framework aims to reduce errors in learning for the CRNet by modeling the standard codec with the ACN and improving the coding efficiency.


\section{Conclusion}
In this paper, we proposed a standard compatible deep neural network-based framework for image compression. 
Within this framework, image compression is performed optimally through the existing off-the-shelf standard codecs, CRNet, and PPNet. The ACN was proposed for optimal learning of the entire network and are designed to imitate the forward degradation processes of existing codecs, such as JPEG and HEVC. Proper training strategies were proposed to minimize errors due to the objective function with approximation. 
The experimental results reveal that this approach outperforms the existing codecs and end-to-end learnable image compression algorithms. 
For future work, we will extend this work to video compression tasks, which are more challenging and complex to model because of the high complexity and temporal dynamics.

\ifCLASSOPTIONcaptionsoff
  \newpage
\fi

\bibliographystyle{IEEEtran}

\bibliography{main}

\begin{IEEEbiography}[{\includegraphics[width=1in,height=1.25in,clip,keepaspectratio]{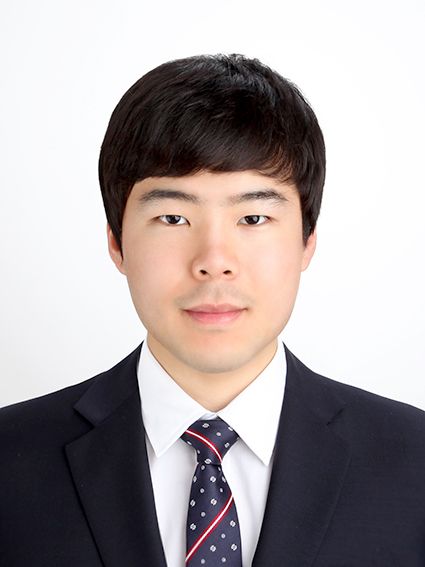}}]{Hanbin Son}
	received the B.S. degree in electrical and electronic engineering from
	Yonsei University, Seoul, Korea, in 2016,
	where he is currently pursuing the Ph.D. degree in electrical and electronic engineering.
	His current research interests include video compression and image processing via deep learning.
\end{IEEEbiography}

\begin{IEEEbiography}[{\includegraphics[width=1in,height=1.25in,clip,keepaspectratio]{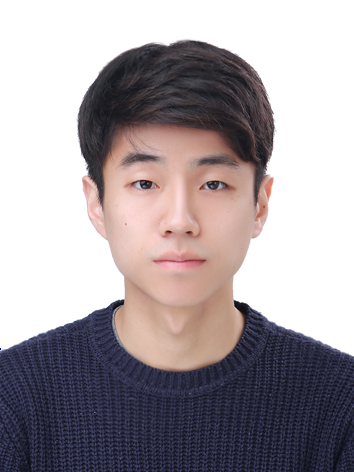}}]{Taeoh Kim}
	received his B.S. degree in Electrical and Electronic Engineering from Yonsei University, Seoul, South Korea, in 2015, in where he is currently pursuing the Ph.D. degree. His current research interests include image/video restoration, face recognition, and video recognition.
\end{IEEEbiography}

\begin{IEEEbiography}[{\includegraphics[width=1in,height=1.25in,clip,keepaspectratio]{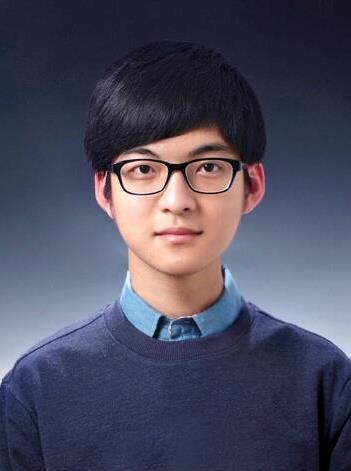}}]{Hyeongmin Lee}
	 is a Ph.D. student of Electrical and Electronic Engineering from Yonsei University, Seoul, South Korea, where he received his B.S. degree in 2018. His research interests include computer vision, computational photography, and video processing.
\end{IEEEbiography}

\begin{IEEEbiography}[{\includegraphics[width=1in,height=1.25in,clip,keepaspectratio]{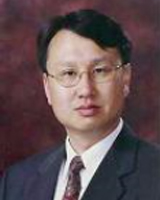}}]{Sangyoun Lee}
	(M’04) received his B.S. and M.S. degrees in Electrical and Electronic Engineering from Yonsei University, Seoul, South Korea, in 1987 and 1989, respectively, and his Ph.D. degree in Electrical and Computer Engineering from the Georgia Institute of Technology, Atlanta, GA, USA in 1999. He is currently a Professor of Electrical and Electronic Engineering with the Graduate School, and the Head of the Image and Video Pattern Recognition Laboratory, Yonsei University. His research interests include all aspects of computer vision, with a special focus on pattern recognition for face detection and recognition, advanced driver-assistance systems, and video codecs.
\end{IEEEbiography}

\end{document}